\documentclass[twocolumn]{aastex63} 

\usepackage{amsmath}
\usepackage{amssymb}
\usepackage{gensymb}
\usepackage{multirow}
\usepackage{booktabs}
\usepackage{subfigure} 

\usepackage{xcolor}

\newcommand{\kms}{km\,s${}^{-1}$}
\newcommand{\pckpc}{pc\,kpc${}^{-1}\:$}
\newcommand{\kmskpc}{km\,s${}^{-1}$\,kpc${}^{-1}\:$}

\newcommand\HI{$\textrm{H\,}\scriptstyle\mathrm{I\:}$}
\newcommand\HII{$\textrm{H\,}\scriptstyle\mathrm{II\:}$}

\received{June 2022}
\revised{July 2022}
\accepted{August 2022}
\shorttitle{Topography of the Solar Neighborhood}
\shortauthors{Alfaro et al.}

\defcitealias{ACCD1991}{ACCD}
\defcitealias{EfremovSitnik88}{ES88}

\begin{document}

\title{Topography of the Young Galactic Disk:
Spatial and Kinematic Patterns of the Clustered Star Formation in the Solar Neighborhood}


\correspondingauthor{Emilio J. Alfaro}
\email{emilio@iaa.es, mjimenez@iaa.es, mcarmen.sanchez@uca.es}

\author[0000-0002-2234-7035]{Emilio J. Alfaro}
\affiliation{Instituto de Astrof\'isica de Andaluc\'ia (CSIC), E-18\,008 Granada, Spain}

\author[0000-0002-8651-9377]{Manuel Jim\'enez}
\affiliation{Instituto de Astrof\'isica de Andaluc\'ia (CSIC), E-18\,008 Granada, Spain}

\author[0000-0003-0829-4157]{M. Carmen S\'anchez-Gil}
\affiliation{Depto. de Estad\'istica e Investigaci\'on Operativa , Universidad de C\'adiz, \\ 
Campus Universitario R\'io San Pedro s/n, E-11\,510 Puerto Real, Spain}

\author[0000-0002-0042-3180]{N\'estor S\'anchez}
\affiliation{Universidad Internacional de Valencia (VIU),
C/Pintor Sorolla 21, E-46\,002 Valencia, Spain}

\author[0000-0002-6823-3869]{Marta González}
\affiliation{Universidad Internacional de Valencia (VIU),
C/Pintor Sorolla 21, E-46\,002 Valencia, Spain}

\author[0000-0003-0825-3443]{Jes\'us Ma\'iz Apell\'aniz}
\affiliation{Centro de Astrobiolog\'ia (CAB), CSIC-INTA, campus ESAC, camino bajo del castillo s.n., E-28\,692 Madrid, Spain.}

\begin{abstract}

The accuracy in determining the spatial-kinematical parameters of open clusters makes them ideal tracers of the Galactic structure. Young open clusters (YOCs) are the main representative of the clustered star formation mode, which identifies how most of the stars in the Galaxy form. We apply the Kriging technique to a sample of \textit{Gaia} YOCs within a 3.5 kpc radius around the Sun and log(age) $\leq$ 7.5, age in years, to obtain the $Z(X,Y)$ and $V_Z(X,Y)$ maps. The previous work by \citet{ACCD1991} showed that Kriging can provide reliable results even with small data samples ($N \sim 100$). We approach the 3D spatial and vertical velocity field structure of the Galactic disk defined by YOCs and analyze the hierarchy of the stellar cluster formation, which shows a rich hierarchical structure, displaying complexes embedded within each other. We discuss the fundamental characteristics of the methodology used to perform the mapping and point out the main results obtained in phenomenological terms.  Both the 3D spatial distribution and the vertical velocity field reveal a complex disk structure with a high degree of substructures. Their analysis provides clues about the main physical mechanisms that shape the phase space of the clustered star formation in this Galactic area. Warp, corrugations, and high local deviations in $Z$ and $V_Z$, appear intimately connected in a single but intricate scenario.

\end{abstract}

\keywords{stellar clusters, star formation --- statistical methods --- catalogs --- surveys --- Milky Way structure}



\section{Introduction}
\label{sec:intro}

The Milky Way disk is highly structured. In two-dimensional (2D) space, the spiral morphology is the most conspicuous feature \citep[][are examples of former views]{becker1964, kerr1970, SK1975}, although its origin and even the number of grand design arms are still a matter of debate \citep[see][and references therein]{Turner2014, Poggio2021, Martinez_Medina2022}. 

Since the late 1950s, we well know that various Galactic disk constituents  have spatial distributions far away from planarity. The first evidence came from the study of the \HI distribution carried out to redefine the fundamentals  of the Galactic coordinate system \citep{kerr1958, gum1960}. Atomic hydrogen exhibited non-planar structures at different spatial scales and Galactic locations. The outermost edge of the disk showed a warped shape that could reach amplitudes of several hundreds parsecs \citep{gum1960, henderson1982}. In the central region, a tilted \HI distribution with respect to the formal plane could be observed \citep{LB1980}, and in the internal Galactic regions some ripples in $Z$ (vertical distance from the formal plane) of low amplitude ($\approx$  50 pc) with variable spatial scales of the order of a few kpc, were the most outstanding features \citep[][]{dixon1967, henderson1967, kerr1970, VQ1970, quiroga1974, lockman1977, QS1977, SF1986}. Similar structures appeared to be also outlined by different molecular species \citep[see][among others]{CD1983, SF1986, KK1988, WBBK1990,combes1991,malhotra1994}. The young and massive stars that delineate the spiral structure in the solar neighborhood provided the first evidences of a stellar population with a vertical distribution far from a planar symmetry. For reasons of proximity, the Local arm (LA) has been the main feature that revealed the complicated morphology shown by the young stellar components of the disk. 
\cite{dixon1967}, using a sample of early B stars, defined  the locus of the LA in the solar neighborhood, and drew, for the first time, its vertical profile (his Fig. 4), concluding that: \textit{Spiral arms are ribbon shaped.}

Later studies kept their focus on the LA, increasing the length of the analyzed segment, and including new spiral arm tracers. Particularly, the innovative \HI analysis performed by Quiroga and collaborators \citep[][]{VQ1970, quiroga1974, QS1977}, not only provided a very detailed description of the wave-like vertical structure of the arms, but also pointed out some possible formation scenarios. Other star formation tracers were soon incorporated to this picture, such as fertile gas (\HI, CO molecular clouds, etc.), young stellar objects with different masses and evolutionary states, as well as typical signposts of the feedback between the newborn objects and the leftover gas \citep[\HII regions, reflection nebulae, etc., see][for a compilation  of different works]{SF1986}. Due to all these efforts, at the end of the 1980s the spatial corrugations associated with the three spiral arms of the solar neighborhood had been analyzed in great detail. Then,  the question arose whether this phenomenology was only associated with the spiral structure or if, on the contrary, it was a phenomenon that encompassed the entire Galactic disk. The only well known extended structure in the solar neighborhood at that time, the Gould Belt (GB), had to be incorporated into this debate.

If we consider a Galactic area within a radius of 1 kpc around the Sun, the typical star formation tracers (originally the brightest stars in the sky) seem to split into two \textit{stellar systems}: one, the Milky Way disk, and the other, tilted around 20$^{\degree}$ with respect to the latter, the GB \citep[][among many others]{Herchell1847, Gould1879, VdB1966, Taylor1987, Poppel1997, Palous_GB1997, Elias2006a}. Although both the Galactic corrugations and the GB rest on, and are described by, the same set of young Population I-type objects, no study had jointly analyzed both phenomenologies. The GB is not considered today to be representative of a single and extended region of coherent star formation \citep{MAF1999, Sanchez2007a, EACC2009}. All the properties and structural parameters estimated for the GB can be obtained considering only two star-forming regions, Orion and Sco-Cen, with  different star formation modes (clustered star formation in Orion, and \textit{looser} star formation pattern in Sco-Cen) \citep{EACC2009}. On the other hand, Sco-Cen's kinematics is not compatible with that expected for a coherent star-forming ring (or disk) surviving longer than a few Ma \citep{MAF1999}. Thus, since 2009 we know that the GB is not a single star-forming complex with a coherent phase-space structure. Quoting \citet{EACC2009}: {\sl "... perhaps we should definitely drop the traditional hypothesis of a single, common origin for all the features of the GB, and begin to look at it as a hazardous alignment -from our point of view- of at least two of the many clumps in the hierarchical structure of the Local Arm, with different densities and star formation history. In this sense, the GB would be simply the projection over the sky of the recent star formation in the clouds close to the Sun but located far away from the fundamental Galactic plane."} This point of view was lately adopted by \citet{Bouy2015}. Nevertheless, even considering that we are not dealing with a coherent star-forming system its apparent morphology must also be explained by the same mechanisms that would generate the observed wobbly disk.

In 1991, a three-dimensional (3D) map of the Galactic disk in the solar neighborhood was obtained for the first time, based on the distribution of young open clusters (YOCs) \citep[][ hereinafter \citetalias{ACCD1991}]{ACCD1991}. The map was tailored using the Kriging technique \citep{Krige1951, Matheron19631246}, which allowed the authors obtaining a detailed 3D view of the Galactic disk with only 82 sample points. It was also the first time that the mining prospecting toolkit, named Kriging or \textit {Krigeage} in its French designation, was incorporated into astronomy \citep{Lombardi2002}. 
Two years later, the idea of a whole corrugated disk rather than a bunch of undulating spiral arms was supported by the 3D analysis of the Galactic Cepheids performed by \cite{berdnikov1993}.

Open clusters have been widely used in the study of the structure of the Galactic disk. Prior to Gaia there have been several catalogs and compilations of these objects whose best known representatives are: \cite{Lynga1982, Janes1982, Dias2002, Kharchenko2005}, and their updates. The determination of the parallaxes and proper motions of 20 stellar clusters obtained with the final Hipparcos calibration \citep{vanLeeuwen2009} deserves special mention and has played an important role as  observational constraint of stellar evolution theories.  Few of the papers derived from these catalogs deal with the 3D structure of the young Galactic disk. The vertical structure of the III Galactic quadrant has been one of the best studied issues \citep[][]{Moitinho2006, Vazquez2008}. Three  observational features were reported in this quadrant: an extensive and deep depression called {\sl Big Dent} \citepalias{ACCD1991}, hints of the southern stellar warp \citep{DIRBE1994} and a stellar overdensity in Canis Major \citep{Newberg2002}. 

The vertical velocity field in the Milky Way has had a less rich history  mainly due to the small number of good quality kinematic data. It was  after  the publication of systematic stellar-radial-velocity catalogs (RAVE, \cite{RAVE2006}; APOGEE, \cite{APOGEE2008}; LAMOST, \cite{LAMOST2012}; GES, \cite{GES2012}; GALAH, \cite{GALAH2015MNRAS}; among others), and \textit{Gaia} releases \citep{GaiaDR12016, GaiaDR2_2018, GaiaEDR32021} that the kinematic analysis could  be extended to significant areas of the Galactic disk. 

\cite{henderson1968}, analyzing the velocity variations of the 21-cm \HI line, claimed the detection of motions departing from circular rotation, which could also be interpreted as variations of the velocity vertical component. These results were subsequently confirmed and refined by other authors \citep[e.g.,][]{YW1973, LB1981, FS1985}. \cite{Comeron1999}  found a systematic variation of the vertical component of the velocity in the region associated with the GB, pointing to an oscillatory motion around an axis nearly parallel to the solar Galactocentric radius. This study was mainly based on OB stars from the \textit{Hipparcos} catalog \citep{Perryman1997}. 

The corrugations of the Galactic disk, together with the warping of the outermost regions, appear to be a universal phenomenon that is observed in most edge-on spiral galaxies \citep[e.g.,][among others]{battaner1990, Sanchez-Saavedra1990, florido1991, florido1992, uson2008}. There is also evidence of the undulating behavior of the vertical velocity component in some external disks \citep{alfaro2001, SGAP2015}. It seems clear that both structures represent projections on different subspaces of the same phenomenon that would affect the entire phase space. However, this connection between spatial structure and velocity field has not been easy to establish due to the lack of quality data for an adequate sample of tracers, either in distance or velocity. Subsequently, they have limited their study to a few local Galactic regions, the GB essentially, before \textit{Gaia} advent.

Detailed analysis of the first topography of the young Galactic disk \citep[\citetalias{ACCD1991},][]{alfaro1992} led these authors to propose that: \textit{The three-dimensional spatial distribution of the Young Population I-type objects can help determine more clearly the structure and size of the young star-gas complexes and give some clues as to the origin and development of these large clouds and to the possible mechanisms driving the large-scale star formation in our Galaxy.} This is the main objective of our work, to determine the spatial structure  and the vertical velocity field in a wide region of the young Galactic disk around the Sun. That is, to establish observational constraints in phase-space subspaces that enable to link the formation of stellar clusters and the generation of  a wobbly disk. Now we can do that because of \textit{Gaia}.

The \textit{Gaia} astrometric mission has revolutionized most astronomical fields and, in particular, our Milky Way view. Several teams have focused on the study of the Galactic open cluster system. They catalogued the actual clusters and measured their main physical variables, including, in some cases,  complete  kinematic information \citep{soubiran2018, Cantat-Gaudin2020, Dias2021, Tarricq2021}. YOCs are the main representative of the clustered star formation mode, which identifies the way in which most of the stars in the Galaxy form \citep{Lada2003, Portegies2010}. Nevertheless, the proportion of stars born in bounded clusters, and how them migrate to the disk  are still a matter of debate \citep[][and references therein]{Bermudas2022, Pang2022, Kroupa2022}.    

We choose Kriging \citep{Matheron19631246} to obtain the functions $ Z(X,Y) $ and $ V_Z(X,Y) $, as the mathematical method best suited to the characteristics of the problem. If we consider the star clusters as probes of the 3D disk structure, we can infer that the observational sample is more similar to what a gold deposit would show than to another ore more reproducible by continuous and derivable functions.  In other words, the distribution of YOCs appears to be a realization of a stochastic process whose analysis requires the right mathematical tools. As has been shown in previous works \citepalias{ACCD1991}, Kriging is able to provide reliable results even working with not too large samples \citep[][N $\approx$ 100]{Oliver201456}.

With this approach, we address the analysis of the following items: 
\begin{enumerate}
    \item The hierarchy of the stellar cluster formation.
    \item The 3D spatial distribution of star formation in clusters and its comparison with other young stellar objects.
    \item The vertical velocity field of the YOCs ($ \mathrm{log} \ t \leq  7.5 $; $ t $ in years) in an area of $\sim\,$40 kpc$^2$  around the Sun. 
    \item The role of the  inner solar neighborhood within the larger spatial area analyzed in this work.
    \item The warp as drawn by the YOCs.
    \item The physical mechanisms shaping the 3D Galactic disk structure.
\end{enumerate}

The paper is organized in six sections, being the first this Introduction. \S{2} describes the data sets and the physical variables that we make use in this work. Kriging technique, and in particular the algorithm employed in our estimations, is described and discussed in \S{3}. \S 4 is devoted to obtain a view of the hierarchical clustering structure and the 3D spatial pattern. In addition, we obtain the vertical velocity field drawn by the YOC population. In \S{5} we discuss our results in relation to the scientific objectives listed in this Introduction, and finally, \S{6} summarizes our conclusions. 

\section{Data} 
\label{sec:data}
\subsection{Input Data}
\label{subsec:input-data}

Stellar open clusters exhibit several advantages as tracers of the Galactic disk structure. Their properties can generally be estimated with greater precision compared to isolated stars \citep{becker1964}. 
The outstanding accuracy in parallax and proper motion measurements achieved by \textit{Gaia} mission has enabled the elaboration of catalogs of such objects in terms of distance and age of a unprecedented quality. We rely on the catalogs published by  \cite{Cantat-Gaudin2020} and  \cite{Tarricq2021}, which primarily leverage \textit{Gaia} data release 2 (DR2), to assemble a set of clusters within a radius $ r \leq 3.5 $ kpc around the Sun and age $ \leq 10^{7.5} $ years.  
Thus, these selection criteria result in a sample of YOCs very representative of the places where they were born.

The catalog presented in \citet{Cantat-Gaudin2020} represents the most complete collection of stellar clusters detected and analyzed from \textit{Gaia}~DR2 individual stars, following a homogeneous procedure in the selection of cluster members\footnote{The compilation of \citet{Dias2021} lists more clusters with kinematic information than \citet{Tarricq2021}. We chose the latter because the cluster members were selected with the same method, in a homogeneous way.}. Both, the five astrometric variables and the three \textit{Gaia} photometric bands are employed to list the probable members and determine the physical properties of the cluster. Their methodology consists in a data-driven approach based on the traditional isochrone fitting of the observational color-magnitude diagram (CMD) by means of an artificial neural network (ANN). This ANN inputs a flatten representation of the cluster's CMD along with two additional quantities estimated from it and the median parallax, and learns the correlation between these inputs and the targets: cluster's age, distance modulus, and extinction. The network is trained on a set of 347 well known clusters that are considered as reference points \citep{Bossini2019}. Then, it is applied to a set of 2017 clusters identified in previous studies \citep{Cantat-Gaudin2020633, Liu2019} using the UPMASK algorithm \citep{KM2014} and involving photometric and astrometric data. At the end of the process, they compile a catalog\footnote{The cluster catalog and complete list of members are available at \url{http://cdsarc.u-strasbg.fr/viz-bin/cat/J/A+A/640/A1}.} of 1867 clusters with reliable parameters, including their position in Galactic Cartesian coordinates, age, parallax, and extinction. However, data about their proper motions are not supplied. 

In contrast, the catalog produced by \citet{Tarricq2021} constitutes the largest collection of open clusters including radial velocity (RV) measurements, as a result of the study carried out to investigate the kinematic behavior of the open cluster population. These RV data are primarily gathered from \textit{Gaia}~DR2, although increasing the number of stars and measures per star by considering nine additional RV surveys \citep{Tarricq2021}. To compute the RVs, a weighted mean of the RV of each star is calculated, regarding the different measures in the catalogs considered and their errors \citep{Soubiran2013,soubiran2018}. The cluster membership is directly taken from the list of 2017 clusters by \citet{Cantat-Gaudin2020} referred above, from which they also get the mean positions, distances, proper motions, 
and age. As a result, the final catalog\footnote{The cluster catalog and complete list of members are available at \url{http://cdsarc.u-strasbg.fr/viz-bin/cat/J/A+A/647/A19}.} includes complete kinematic information for 1382 open clusters, signifying a great improvement with respect to the largest previous compilations \citep{Kharchenko2013,Dias2002}.
 
\begin{figure*}[ht]
\begin{center}
\includegraphics[width=\textwidth]{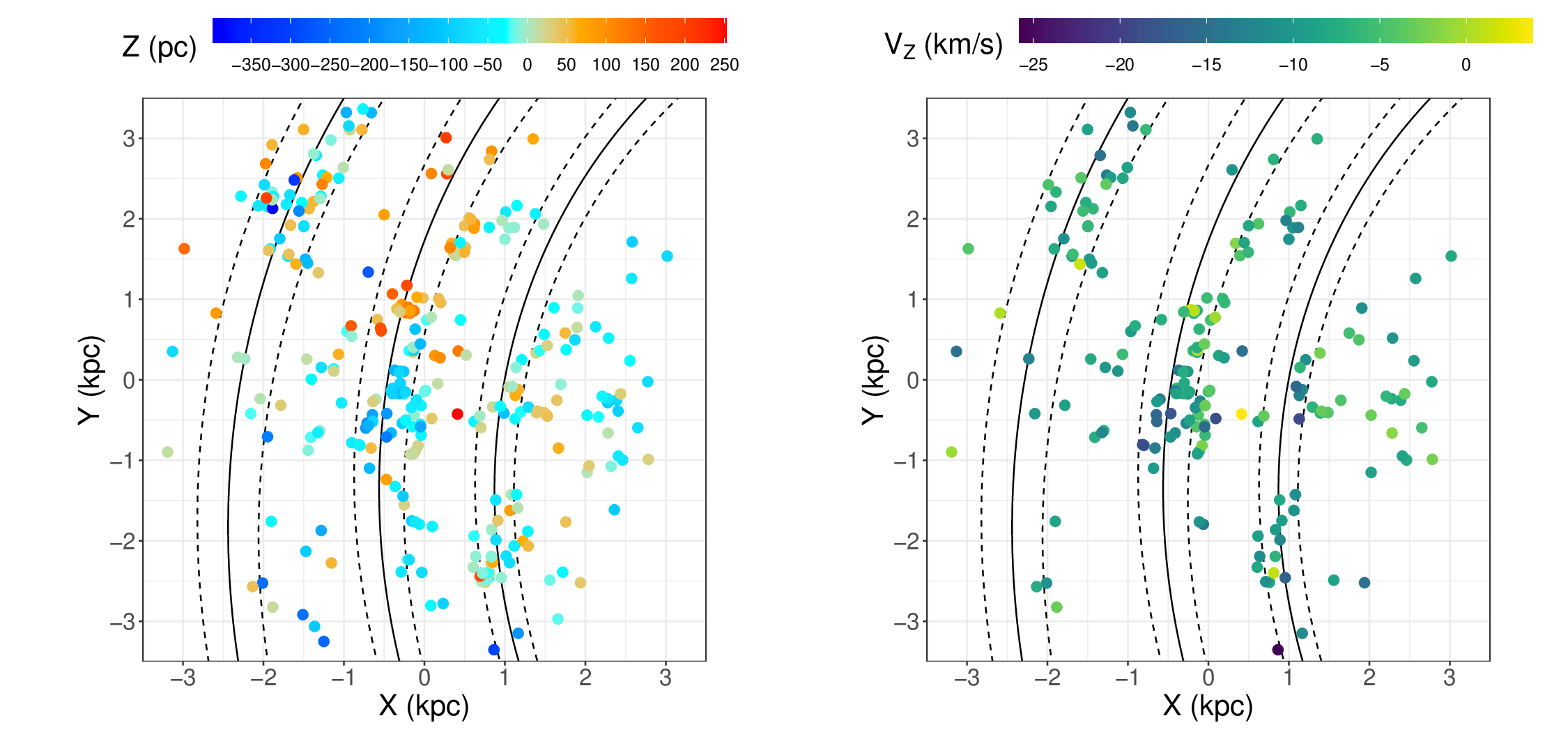}
\caption{G-YOC sample from the merging of \citet{Cantat-Gaudin2020} and \citet{Tarricq2021} catalogs in Galactic Cartesian coordinates. The selection criteria are: $|r| \leq 3.5$ kpc and $ \mathrm{log}(Age) \leq 7.5 $, resulting in 309 YOCs (left) of which 183 have their vertical velocity component measured (right). The color codes for $Z$ (pc) and $V_Z$ (\kms) are shown at the top. The $ X $ and $ Y $ axes take positive values toward the Galactic center and Galactic rotation directions, respectively. The Sun is at $ (0,0) $. The shown spiral arms are, from left to right, Perseus, Local, and Sagittarius. Spiral arms parameters are taken from \citet{2021A&A...652A.162C}, with fixed arms' width based on \cite{2014ApJ...783..130R}.}
\label{fig:Young}
\end{center}
\end{figure*}

We perform several steps starting from these two collections to compose the final sample of YOCs for our study. First, we apply the criteria detailed above about distance and age separately, yielding a total of 301 and 166 YOCs from Cantat-Gaudin and Tarricq catalogs, respectively. Then, we cross-match these two samples, incorporating to the Tarricq sample a set of 143 clusters from Cantat-Gaudin not present in the first one. Of these, 108 have velocity measures that are obtained from the catalog\footnote{This catalog is available at \url{http://cdsarc.u-strasbg.fr/viz-bin/qcat?J/A+A/619/A155}.} provided in \citet{soubiran2018}. This results in a final sample of 309 YOCs with positions and age of which 183 have also vertical velocity component\footnote{There exist overlappings of 158 and 91 clusters in the two cross-matches performed concerning positions and vertical velocities, respectively.}. Figure \ref{fig:Young} displays our final YOC sample on the Galactic plane. 
From now on, we will refer to it as \textit{Gaia}-YOC (G-YOC) sample.

\subsection{Our Reference System}
\label{subsec:Reference}

We choose the formal Galactic plane as the main reference plane. The Cartesian space and velocity coordinates are referenced to the Sun, such that $\Vec{r}_{\odot}\equiv (X\odot,Y\odot,Z\odot)\, =(0,0,0)$ pc and $\Vec{V}_{\odot} \equiv(V_{\odot X},V_{\odot Y}, V_{\odot Z})=(0,0,0)$ \kms. We do not include any additional information to determine the maps $ Z(X,Y) $ and $ V_Z(X,Y) $.  
The reason for this decision is that we do not need to estimate additional constants such as $R_{\odot}$, $Z_{\odot}$, or the solar motion, whose determination would depend in each case on the chosen age group, the area of the Galactic plane studied, and the quality of the data used. This is an inescapable fact if, as has already been shown in different studies, the Galactic disk is a non-stationary system in phase space \citep[e.g.,][]{Widrow2012, 2018Natur.561..360A, Bennet_Bovy2019}. Therefore, all the information gathered from other published works and catalogs are transformed to Galactic coordinates with the Sun as origin of the system, if they were formerly referred to other Galactic reference centers or to another set of coordinates.

We also select other reference lines in the $ (X,Y) $ plane that represent typical features of the Galactic disk, such as spiral arms, and other peculiar loci previously detected in the region such as the Cepheus spur introduced in \citet{Pantaleoni-Gonzalez2021}. There is a great variety of definitions (different formula's parameters, and even different mathematical functions) for the three main spiral arms within the analyzed area. These differences are due to the variety of spiral tracers and the different distance-estimation methods used \citep[e.g.,][and references therein]{Hou2021, xu2021, Poggio2021, Vallee_LA_2022}. In this work, we choose the spiral arms defined by \cite{2021A&A...652A.162C} from the joint population of YOCs and maser tracers. The Cepheus spur has been fitted by us as a linear segment according to the previous information available (Figure 7 in \citet{Pantaleoni-Gonzalez2021}). The spiral arms' equation, as well as the range of values where they and the Cepheus spur are defined, are shown in Equation \ref{eq:logarithmic-spiral} and Table \ref{tab:spiral-arms-params}, respectively. In this logarithmic model, $ R $ and $ \Phi $ represent the Galactocentric radius and azimuth along the arm, respectively, $ (R_{ref}, \Phi_{ref}) $ are the coordinates of a reference point, and $ \psi $ is the pitch angle of the arm. We consider a constant width of the spiral arms for their graphical representation, although there is evidence that the width varies along the arm \citep[e.g.][]{Vallee_anchura_2021}.
\begin{equation}
\label{eq:logarithmic-spiral}
\mathrm{ln} \bigg[ \dfrac{R}{R_{ref}} \bigg] = - \, (\Phi - \Phi_{ref}) \, \mathrm{tan} \psi \;
\end{equation}

\begin{table}[htbp]
\centering
\begin{tabular}{lcccc}
\hline
\multicolumn{5}{c}{\textbf{Spiral arms}} \\\hline
Arm & $ R_{ref} $ (kpc) & $ \Phi_{ref} $ (deg) & $ \psi $ (deg) & Width (kpc) \\\hline
Perseus       & 10.88 & -13.0 & 9.8 & 0.38 \\
Local         & 8.69  & -2.3  & 8.9 & 0.33 \\
Sagittarius   & 7.10  & 3.5   & 10.6 & 0.26 \\\hline\hline
\multicolumn{5}{c}{\textbf{Cepheus spur}} \\\hline
$ X_{min} $ (pc) & $ X_{max} $ (pc) & \multicolumn{3}{c}{Equation (pc)} \\\hline
-2157.2 & 202.0 & \multicolumn{3}{c}{$ Y = 0.6528 X + 995.1418 $} \\\hline 
\end{tabular}
\caption{Spiral arms' parameters according to \citet{2021A&A...652A.162C} and Cepheus spur linear fit based on the study presented in \citet{Pantaleoni-Gonzalez2021}.} \label{tab:spiral-arms-params}
\end{table}

These 2D lines are mere references to be able to compare our results with previous works, but we are far from considering that the spatial and kinematic structure of the young Galactic disk is limited to a few linear features on the plane. We are faced with a 3D structure far from planarity that affects the entire disk, regardless of the stellar population we are analyzing \citep{Widrow2012}. On the other hand, the very definition of Galactic spiral arms, their nature and origin, are not yet fully understood \citep{Turner2014, Lepine2018, Poggio2021,  Vallee_LA_2022, Martinez_Medina2022}. The characterization of spiral arms in the vicinity of the Sun does not fall within the scope of this paper.  However, we will use the main 2D features provided by other authors to highlight some aspects of the intricate 3D pattern of the young Galactic disk, and to analyze its possible origins.

\section{Methods}
\label{sec:methods}

\subsection{Kriging Fundamentals}
\label{subsec:Kriging}

The genesis of Kriging goes back to the mid 20\textsuperscript{th} century, when Danie G. Krige aimed at predicting the location of ore grades considering a spatially correlated data sample in the gold mines of South Africa \citep{Krige1951}. Later in the 1960s, George Matheron developed the theoretical foundations for the method \citep{Matheron19631246}, and since then, it has been widely implemented in many fields such as engineering, mining, geology, meteorology or remote sensing, to name just a few  \citep{Simpson2001129,ChilesDelfiner2000}. Now Kriging is recognized as an extensive family of estimation methods for multi-dimensional stochastic processes \citep{Webster2007}, although it has barely been exploited in astronomy \citep{Pastorello2014}. It provides the best linear unbiased estimator (BLUE), making an optimal use of existing knowledge by exploring the way that a target variable varies in space through a variogram model \citep{Oliver201456}. 
Kriging technique is closely connected with the Gaussian processes regression (GPs), at least in the final objectives. Being the latter more extended in astronomy \citep[e.g.][]{10.7551/mitpress/3206.001.0001,2022arXiv220604627N,2022arXiv220703492W}. Kriging and GPs are families of interpolation methods that have similarities and subtle differences.  
Mathematically, GPs can be considered a generalization of Kriging into higher dimensional space, and, in some cases, both methods can provide a quite similar mathematical solution \citep[see][for a deeper discussion of both techniques]{feigelson_babu_2012, 
Cui_Tao_Pagendam_21}.

Let $ z(\mathbf{r}) $ be a real function defined in some region of the plane $ \mathbb{R}^{2} $ with known values in a certain number of points $ \mathbf{r}_{i}(x_{i}, y_{i}) $ of a sample set $ S $, which is regarded as a single realization of a random process $ Z(\mathbf{r}) $. The basic problem is to achieve a valid approximation $ z(\mathbf{r}) $ of the function $ Z(\mathbf{r}) $ for points not belonging to $ S $, assuming that some sort of spatial correlation underlies for points in $ S $. In its simpler form, a Kriging estimate is a weighted average of the sample data in their neighborhood. However, more elaborated versions have been devised to take into account general trends in $ S $ or several limitations in its knowledge. In addition, a distinctive characteristic of Kriging is that it enables to quantify the goodness of the prediction, providing a variance $ \sigma^2_{Z} $ associated with the estimated function $ z(\mathbf{r}) $. For our purposes here, $ S $ will be the G-YOC sample introduced above, whose $ (X,Y,Z) $ and $ (V_Z,X,Y) $ coordinates are well determined. The goal is to obtain an estimation of $ Z(X,Y) $ and $ V_Z(X,Y) $ on a rectangular grid that covers the area investigated in order to outline the 3D structure of the solar neighborhood and the vertical velocity field.

Under the assumption of stationarity, we can decompose $ Z(\mathbf{r}) = m + e(\mathbf{r}) $, where $ m $ is the mean of the process, $ m = \text{E}[Z(\mathbf{r})] $, and $ e(\mathbf{r}) $ is a zero mean random process whose covariance $ C(\mathbf{h}) = \text{E}[e(\mathbf{r})e(\mathbf{r} + \mathbf{h})] $ is known, with $ \text{E} $ the expectation and $ \mathbf{h}(\mathbf{r}_{i}, \mathbf{r}_{j}) $ the vector separation between pairs of sample points. Due to the aforementioned stationarity, $ m $ does not depend on $ \mathbf{r} $ and $ C(\mathbf{h}) = C(h) $ only depends on the distance (isotropy). However, a weaker hypothesis originally introduced by Matheron is needed in practical situations, known as the intrinsic model \citep{Matheron1965}. By this, $ \text{E}[Z(\mathbf{r} + \mathbf{h}) - Z(\mathbf{r})] = 0 $ and the variance yields $ \text{Var}[Z(\mathbf{r} + \mathbf{h}) - Z(\mathbf{r})] = 2\gamma(h) $. The resulting  $ \gamma(h) $ is the variogram of the random process, which is assumed to measure the spatial correlation in the actual realization of $ Z(\mathbf{r})$. Generally, it is estimated using the method of moments \citep{Matheron1965} as: 
\begin{equation}
\label{eq:variogram}
\hat \gamma(\mathbf{h}) = \frac{1}{2p(\mathbf{h})} \displaystyle \sum_{j = 1}^{p(\mathbf{h})} \Big\{ Z(\mathbf{r}_{j}) - Z(\mathbf{r}_{j} + \mathbf{h}) \Big\}^{2} \;  ,
\end{equation}
\noindent being $ p(\mathbf{h}) $ the number of point pairs separated by a distance $ \mathbf{h} \in \mathbb{R}^{3} $, or $ h \in \mathbb{R} $ in the particular case under isotropy. 
When $ S $ shows a general trend, that is, a gradual variation in space, the method can be generalized considering that $ m = m(\mathbf{r}) $, responsible for the trend. This is the so-called Universal Kriging (UK). This case, $ m = \text{E}[Z(\mathbf{r})] $ is a functional drift and $ e(\mathbf{r}) $ remains as an intrinsic zero mean random process. This is, in the UK model such a trend $Z(r) = m(r)+e(r)$ is modelled as a linear function
\begin{equation}
\label{eq:krig_model}
Z(r) = \sum_{j=0}^p f_j(r)\beta_j + e(r) = {F\beta} + e(r)
\end{equation}
based on $p$ known predictors or spatial regressors $f_j(r)$ evaluated at each observation $r_i$ (also known as design matrix ${F} = (f_1(r),\ldots,f_p(r))$, with $f_j(r) = (f_j(r_1),\ldots,f_j(r_n)$), and $p$ unknown regression coefficients $\mathbf{\beta} = (\beta_1,\ldots,\beta_p)^T$. 
The most common strategy to deal with it is to first estimate the drift by fitting a polynomial curve and then obtaining the residuals $ e(\mathbf{r'}) $ by subtracting the curve values from the sample data. Given that $ e(\mathbf{r'}) $ is expected to be an intrinsic process, the usual Kriging estimator is calculated for the residuals.   

\subsection{The Variogram Estimation}
\label{subsec:var-estimation}

The variogram analysis provides a useful tool for summarizing spatial data and measuring spatial dependence. Moreover, its main contribution is related to the statistical inference of spatially correlated random variables to estimate the value of the spatial variable at an unsampled location. 
The Kriging approach differs from classical regression in that local features can affect the solution. For that reason, a Kriging estimate consider a weighted mean, bearing in mind that some measurements in the vicinity of the location investigated are more closely related to the actual unknown value than further ones. 
Variogram estimation plays a decisive role in the Kriging approach, commonly used to find the optimal values of the weights. 

The variogram estimation splits into two stages:   
(i) the computation of the experimental variogram from the sample set $S$ \citep{Oliver201456},  so-called {\it empirical variogram estimation}, with the subsequent  fitting with some theoretical model $\gamma(h)$; (ii) the application of the previous variogram estimator to the data to make {\it predictions}.

Under the intrinsic model, see Equation \ref{eq:variogram}, the distance $ h $ is split in a certain number of bins resulting in an ordered set of $(h, \gamma(h))$ points. However, the underlying model is assumed to be continuous in all $\mathbf{h}$. 
We get then a variogram estimator by fitting the empirical variogram by some theoretical model. We select the theoretical smooth curve that best fits the experimental variogram in this stage. This function must belong to a family of valid variograms and capture the available data's underlying spatial dependence.
More details on the variogram models and variogram fitting, together with the theoretical model selection by means of cross-validation comparison and the statistics WSSE, the weighted sum of squared errors of the fitted model, can be found at Appendix \ref{App:VMS}.

Figure \ref{fig:variograms_all} shows the variogram fitting of the spatial $Z$ component for G-YOC sample. Black dots represent the experimental variogram. 
We test four theoretical variogram models $\gamma(h)$: the Spherical \eqref{Eq:Sph_VM} and Exponential \eqref{Eq:Exp_VM} variogram models, which are the most widely used, along with the Gaussian \eqref{Eq:Gau_VM} and the Wave \eqref{Eq:Wave_VM} models. The latter was also considered because of its atypical irregular behaviour, taking into account the expected corrugated nature of the $Z$ variable. 
We chose the Spherical variogram model for both $Z$ and $V_Z$ variables. 

Once we get the variogram estimator, which determines the spatial dependency or autocorrelation in our data, we can estimate  $ Z(X,Y) $ and $ V_Z(X,Y) $ functions using the fitted model. 
Therefore, the last stage of the Kriging approach is to use the data to make predictions and create a continuous surface of the phenomenon. Predictions are made at each location of a spatial grid around the studied area, based on the fitted semivariogram model and the spatial arrangement of the closer input data.

\section{Results} 
\label{sec:results}

In this section, we present the results from the Kriging analysis of the 4D phase-space subspace, inferring both the 3D spatial distribution and the vertical velocity field of the young Galactic disk. 
We previously perform a study of the hierarchical spatial distribution of the G-YOC sample.

\subsection{Clustering of Clusters}
\label{subsec:clustering}

Star formation tends to be distributed in space forming groups that in turn contain other smaller groupings, in a cascade of sizes and ages \citep{EfremovElmegreen1998, Elmegreen2000, 2011EAS....51...31E}. In this section we analyze the spatial hierarchy in the distribution of star clusters. To do so, we obtain the fractal dimension in $ (X,Y) $ 2D subspace and perform a clustering analysis of the 3D spatial coordinates.

\subsubsection{Fractal Dimension of the Distribution of \textit{Gaia} YOCs.}

\begin{figure}[htp]
    \centering
    \includegraphics[width = 0.8\columnwidth]{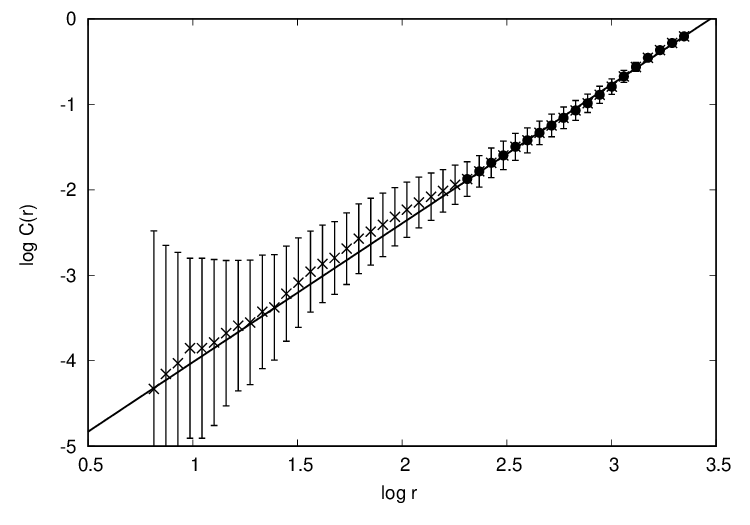}
    \caption{Calculated correlation integral $C(r)$ for the data used in this work. Error bars represent standard deviations $\sigma_C$. Solid line is the best linear fit in the range of reliable values indicated by solid circles, i.e. the range for which $\sigma_C \lesssim C(r)$.}
    \label{fig_Dc}
\end{figure}

One way to characterize the spatial distribution of YOCs and associations is through the fractal dimension, which gives a simple and objective measurement of the degree of clumpiness of the distribution. We calculated the so-called correlation dimension \citep{Grassberger1983} using a previously developed algorithm \citep{Sanchez2007a} that avoids boundary effects and finite-data problems at small scales and, consequently, yields unbiased results even for relatively small sample sizes. The correlation dimension is derived by calculating the correlation integral $C(r)$, that is, the probability of finding a point within a circle of radius $r$ centered on another point. If the points are distributed obeying a fractal geometry of dimension $D_c$, then the relation $C(r) \sim r^{D_c}$ holds and $D_c$ can be estimated as the slope of the best linear fit in a log-log plot. Figure~\ref{fig_Dc} shows the obtained result. The two-dimensional correlation dimension for the sample gives $D_c = 1.62 \pm 0.05$, where uncertainties are estimated from bootstrapping. This fractal behavior extends over a range of scales from $200$ to $2500$ pc and the slope changes observed below $r \sim 200$ pc can be attributed to random fluctuations in $C(r)$ and not to a real transition between two different physical regimes, such as has been reported for the distribution of young stars, YOCs and \HII regions in other galaxies \citep{Odekon2008, Sanchez2010, Menon2021}. 

The degree of clustering of the distribution of star-forming regions may vary among galaxies. Fractal dimensions in the range $1.5 \lesssim D_c \lesssim 2.0$ have been obtained for the distribution of \HII regions in disk galaxies \citep{Sanchez2008}, depending on properties such as brightness, size or age. These distributions are consequence of physical processes acting at large scales, that is, Galactic motions on the disk. At smaller scales where our analysis is relevant, turbulence is probably the main mechanism controlling the underlying structure where coherent star formation occurs. Our result is consistent with a recent detailed analysis of the distribution of $\sim 2500$ young stellar structures in the Large Magellanic Cloud by \citet{Miller2022}, who derived a two-dimensional fractal dimension of around $1.5-1.6$ in the range of spatial scales from 10 pc to 700 pc. The value $D_c = 1.62 \pm 0.05$ implies that the corresponding three-dimensional fractal dimension should be $D_f = 2.5-2.6$ \citep[see Table~1 in][]{Sanchez2008}, which is similar to the range of values $D_f \sim 2.5-2.7$ obtained from emission maps of several spectral lines for different molecular clouds in the Milky Way \citep{Sanchez2005,Sanchez2007b} and for early-type stars in the GB \citep{Sanchez2007a}. In addition, this value is in agreement with simulations of compressively driven supersonic turbulence reported by \citet{Federrath2009}. Overall, the obtained result supports the idea that young, newborn stars and clusters should, on average, reflect the same spatial structure of the interstellar medium from which they were formed.

The fractal structure can be observed over a range of spatial scales between $200$ and $2500$\, pc. We analyze below the hierarchical groupings of the 3D G-YOC sample.

\subsubsection{Clustered Star Formation Hierarchy:
Complexes and Supercomplexes}

We use the density-based tool OPTICS  \citep[\textit{Ordering Points To Identify the Clustering Structure},][]{Ankerstetal99} to analyze the hierarchical spatial substructure of the 3D Galactic Cartesian coordinates $ (X,Y,Z) $ of our sample. Despite the fact that it can be used to retrieve clustering, and has typically been used that way in astrophysics  \citep{Costadoetal17, Canovasetal19,vanTerwisga22}, we want to highlight that, strictly speaking, OPTICS is not a clustering tool, but a density based reordering algorithm. 
OPTICS can be considered an extension of DBSCAN  \citep[\textit{Density Based Spatial Clustering of Applications with Noise},][]{Esteretal96}. We briefly explain the main characteristics of both algorithms, and refer the reader to \citet{Gonzalezetal21} for a more detailed  but intuitive explanation of how DBSCAN works, to \citet{Canovasetal19} for a global comparison of density-based clustering algorithms in an astrophysical context, and to the original papers by \citet{Esteretal96} and \citet{Ankerstetal99} for a complete and rigorous description of both algorithms. In essence, DBSCAN decides whether a point belongs to a cluster or to noise based on whether a minimum number $N_{min}$ of neighbors is within a radius  of size $\epsilon$ around that point. The parameters $N_{min}$ and $\epsilon$ define a typical scale and density of the clusters retrieved by the method, which limits the analysis when considering samples where structure is present at several scales and densities. OPTICS overcomes this limitation by defining the ordering of the points based on the concept of reachability distance between points. For a fixed $N_{min}$, this distance can be intuitively understood as the $\epsilon_{N_{min}}$ needed for both points to be in the same DBSCAN cluster. The main product of the algorithm is the reachability plot, where the reachability distance of each point is shown against the ordering index provided by OPTICS. The reachability plot gives a complete and intuitive visual description of the clustering structure of a sample, since clusters and dense subsets are seen as dents or valleys in the plot. 

\begin{table*}[htp]
\centering
\caption{Central Cartesian coordinates and 3D-diameters of the hierarchical clumps retrieved by OPTICS. Cluster 1, of level 3, which encompasses almost the entire sample is not shown. Correspondences between the stellar complexes and supercomplexes listed by \citetalias{EfremovSitnik88} and those detected in this work are detailed in the last column. Each separate hierarchical family is identified with a capital letter, in accordance to Figures \ref{FigOpticsGral} and \ref{FigCompareEfremov}. N is the number of YOCs in each group. The rest of names are self-explanatory.}
\label{tableOptics}
\begin{tabular}{ccccccccccc}
\hline
Family &Id &  N    &Level&  X (pc)  & Y (pc) &Z (pc) & 3D diam. (pc) & ES88 Name\\
\hline
 \hline
A&2 &  85     &2& -250.6&	42.7&	-22.2&2338.5 & Local, 4 (10,8) \\
&3 &  20   &0& -157.5&	899.4&	72.1&626.0 & \\
&4 &  34    &1& -201.9&	-242.2&	-70.8&993.0 & 4 \\
&5 & 16&     0 & -319.5&	-109.7&	-98.9&  294.6 & 4 \\
&6  &   7  &   0  &-720.6&	-611.4&	-120.7& 425.0 & \\
\hline
B&7 &35   &  1  &1394.7&40.7&	-5.4& 2038.2 & 1-I, 2-I \\
&8 & 9    & 0  &1673.4&	484.4&	-20.5& 1006.2 & 2-I \\
\hline
C&9 & 17   & 0 &2349.4&	-585.8&	-36.8&  1357.5 & 17 \\
\hline
D&10   &  25   &  0& 895.3&	-2207.3&	1.4& 1019.& 15-IV, 16 \\
\hline
E&11&     9    & 0  &415.4&	1651.8&	36.2& 390.2 & 3 \\
&12&     22    & 1  &730.5&	1833.2&	17.6& 1205.4 & 3-3* \\
\hline
F&13 &48 &1    &-1510.7&	2326.8&	-30.8& 2142.3 & 7-II, 6-II, 5 \\
&14 &6  &0&   -839.8	&3228.0&	-52.5&  373.8 & \\
  \hline
G&15 &8  &0&   -1691.7&	-2630.3&	-112.6&  1642.3 & 13* \\\hline
\end{tabular}
\end{table*}

This view of clusters as valleys in the reachability plot allows for retrieval of the clusters in the sample: if a specific reachability distance $\epsilon$ is fixed, the DBSCAN clusters with $N_{min}$ and $\epsilon$ can be directly obtained from the plot as the valleys deeper than $\epsilon$, i.e. characterized by reachability distances smaller than $\epsilon$. The original paper by \citet{Ankerstetal99} proposes a different possibility for cluster retrieval where the clusters are separated based on the steepness of the slope of the valleys formed in the reachability plot instead of their depth, fixing a parameter $\xi \in (0,1)$ associated to the difference in density between adjacent points. As explained in the original paper, with this procedure clusters are features in the reachability plot that start with a downward slope, end with an upward slope and fulfill requirements for both a minimum number of points and a minimum density increment with respect to its environment.

\begin{figure*}[ht!]
\centering
\includegraphics[width=0.8 \textwidth]{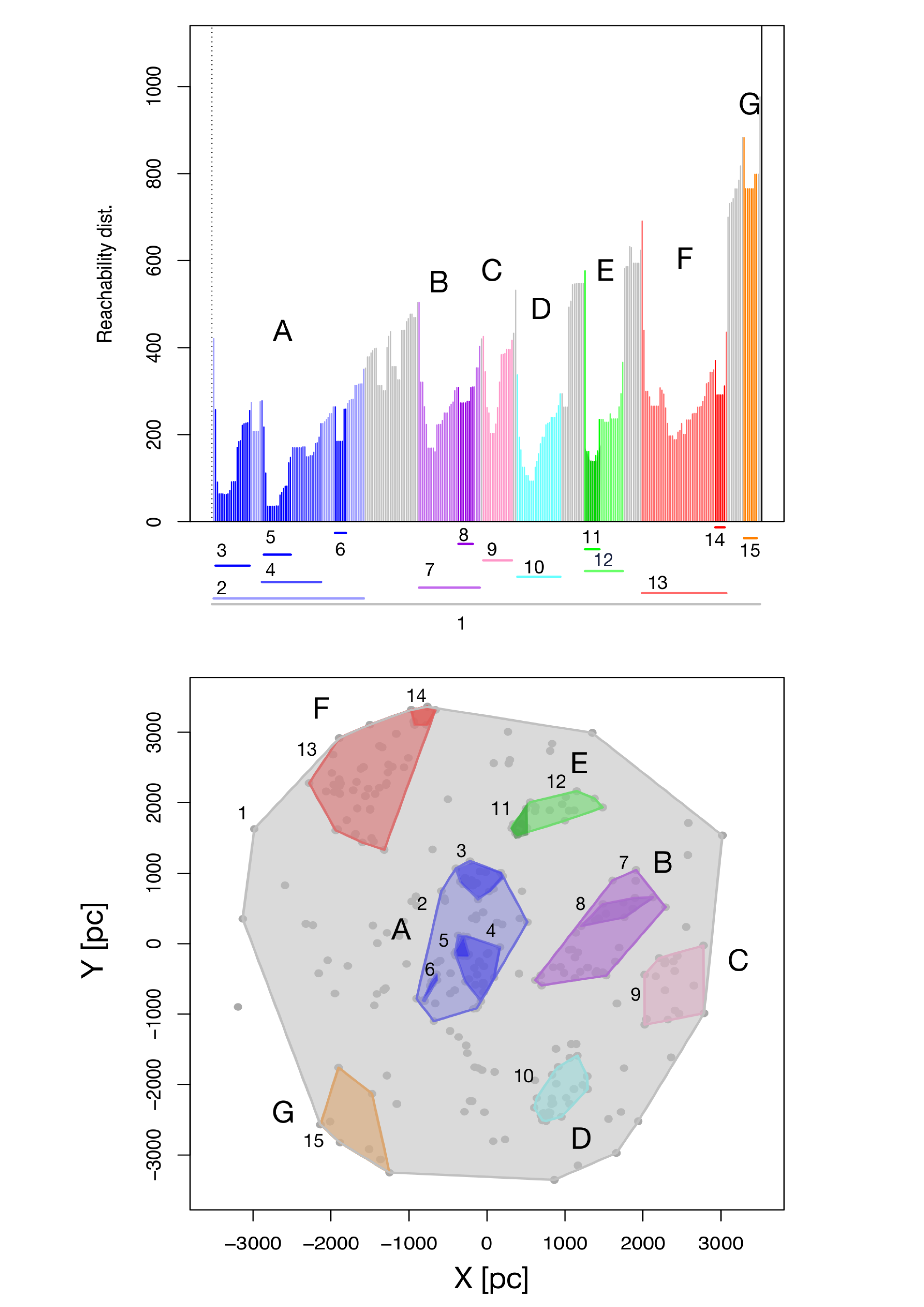}
\caption{Top: Reachability plots obtained by OPTICS. 
Bottom: Projected convex hulls of the structures obtained onto the Galactic plane. Each separate complex/supercomplex is shown in a different color, and tagged according to Table \ref{tableOptics}.} 
\label{FigOpticsGral}
\end{figure*}

In this work, we use the $\xi$ strategy since it takes advantage of the multi-scale approach of OPTICS, allowing for hierarchical structures to emerge, with groupings potentially embedded within each other. 
The main results of our analysis are shown in Figure  \ref{FigOpticsGral} and Table \ref{tableOptics}, and reveal the clustering hierarchy of YOCs. If we define the hierarchical level of a cluster recursively, with level 0 clusters not containing any other cluster, and clusters of level $l$ containing clusters of level $l-1$, the clustering obtained in this work  reaches  up to level 3, corresponding to the entire distribution.
We have fixed $N_{min}=6$ as a reasonable balance between noise and the detection of small structures, and have explored the parameter $\xi$, choosing $\xi=0.09$, which prioritizes the significance of the structures detected instead of richness of hierarchy. Additionally, for further confirmation of the robustness of the retrieval, we have also run OPTICS with $N_{min}=8$ and confirmed that all the main families in Table \ref{tableOptics} are also present in the corresponding reachability plot, while other dents (like those between families A and B in the upper plot of Figure \ref{FigOpticsGral}) are smoothed out.

Top panel in Figure \ref{FigOpticsGral} shows the reachability plot for G-YOC sample. 
That plot shows numerous valleys, some embedded within each other, revealing the presence of a hierarchical structure. The bottom panel of Figure \ref{FigOpticsGral} shows the hierarchy of groupings found in the Galactic plane, plotting only the $(X,Y)$ variables, although the hierarchical distribution is based on three spatial Cartesian coordinates. 
The results from OPTICS show that the G-YOC population is, as already pointed by the fractal analysis in the previous section, structured at various densities and scales, showing both global and local hierarchical clustering all over the region sampled.

\begin{figure*}[htbp]
\begin{center}
\includegraphics[width=\textwidth]{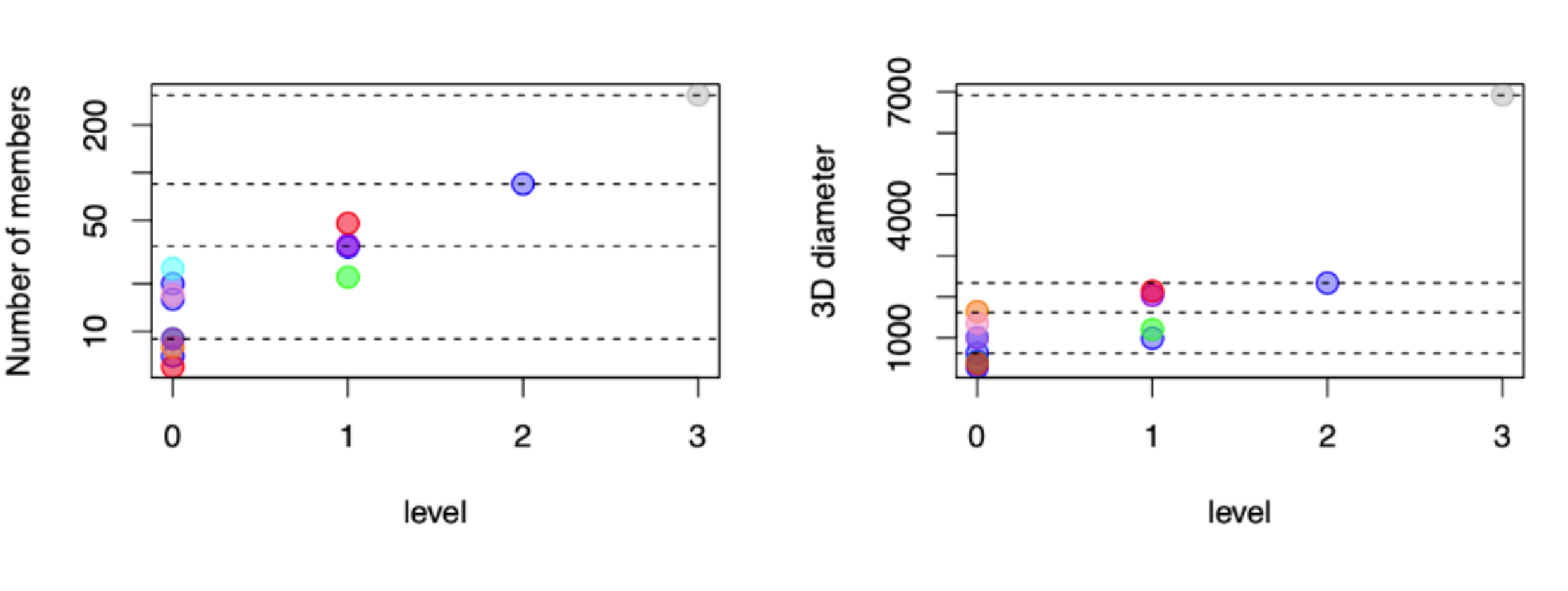}
\caption{Left: Number of YOCs in each structure, as a function of their hierarchical level. Right: 3D diameter (pc) of the structures found in this work as a function of their hierarchical level. In both panels, the color code indicates the hierarchical family of the structure, and the horizontal dashed lines correspond to the median value for each level.}
\label{FigSizeStructures}
\end{center}
\end{figure*}

In Figure \ref{FigSizeStructures} we explore some of the characteristics of the groupings, in relation to their hierarchical level. The top panel shows the size (measured as its tree-dimensional diameter) of each grouping against its level. All structures below level 2 are well within the fractal spatial scale of 200-2500 pc. We remind the reader that the only structure of level 3 is almost the complete sample, and thus has a diameter corresponding to the 3.5 kpc initial radius. We observe a clear increasing trend in median size with the hierarchical level.  A similar trend can be found in the population of the groupings, shown in the lower panel of Figure \ref{FigSizeStructures}.

Some aspects of interest should be noted: (a) the main groupings coincide with the overdensities of massive OB stars  (sample M)  \citep{Pantaleoni-Gonzalez2021} and with those detected from  Upper Main Sequence (UMS) stars, young Galactic clusters and Cepheid stars younger than 100 Ma \citep{Poggio2021} in the common area analyzed, and (b) there is full agreement between the hierarchical stellar clumps classified in this work and the stellar complexes and supercomplexes detected  from the 2D distribution of YOCs, OB associations and \HII regions by \citet[][ hereinafter \citetalias{EfremovSitnik88}]{EfremovSitnik88}. 

If we compare the position and size of the supercomplexes obtained from our analysis with the distribution of the  young stars' filaments  from \citet{Kounkel2020} (top panel of their Fig. 1, corresponding to $ 7.0 < \mathrm{log}(age) < 7.7 $), we see many similarities and some differences. The supercomplex A is practically the LA central grouping and its $spine$, defined in \citet{Kounkel2020} as the locus perpendicular to adjacent filaments, connects the lower hierarchy complexes within A. YOCs, and looser and extended concentrations of young stars, show a similar pattern in the LA. 
The same happens in the Sagittarius region, and in the area associated with the Scutum arm, although the groupings (B and C) of the G-YOC sample, appear better separated than the structures of \citet{Kounkel2020} in the same area.  In contrast, the lower segment of the Perseus arm, in the III Galactic quadrant, shows to be richer in filaments and other looser OB groupings  than in YOCs.
The Cygnus region, on the other hand, defined by complex E, shows a high concentration of clusters, as does Carina too (complex D).
Fig. 1 (top panel) in \citet{Kounkel2020} encompasses a slightly wider range of ages than our Figure \ref{FigOpticsGral} (bottom panel). In addition, our sample is representative of the current clustered star formation. This last issue could be the cause for the main differences found between both distributions. Star formation in Cygnus and Carina appears to occur mainly in clusters, while few young and massive stars are isolated or located in looser groups.

\begin{figure*}[htbp]
\begin{center}
\includegraphics[width=0.8\textwidth]{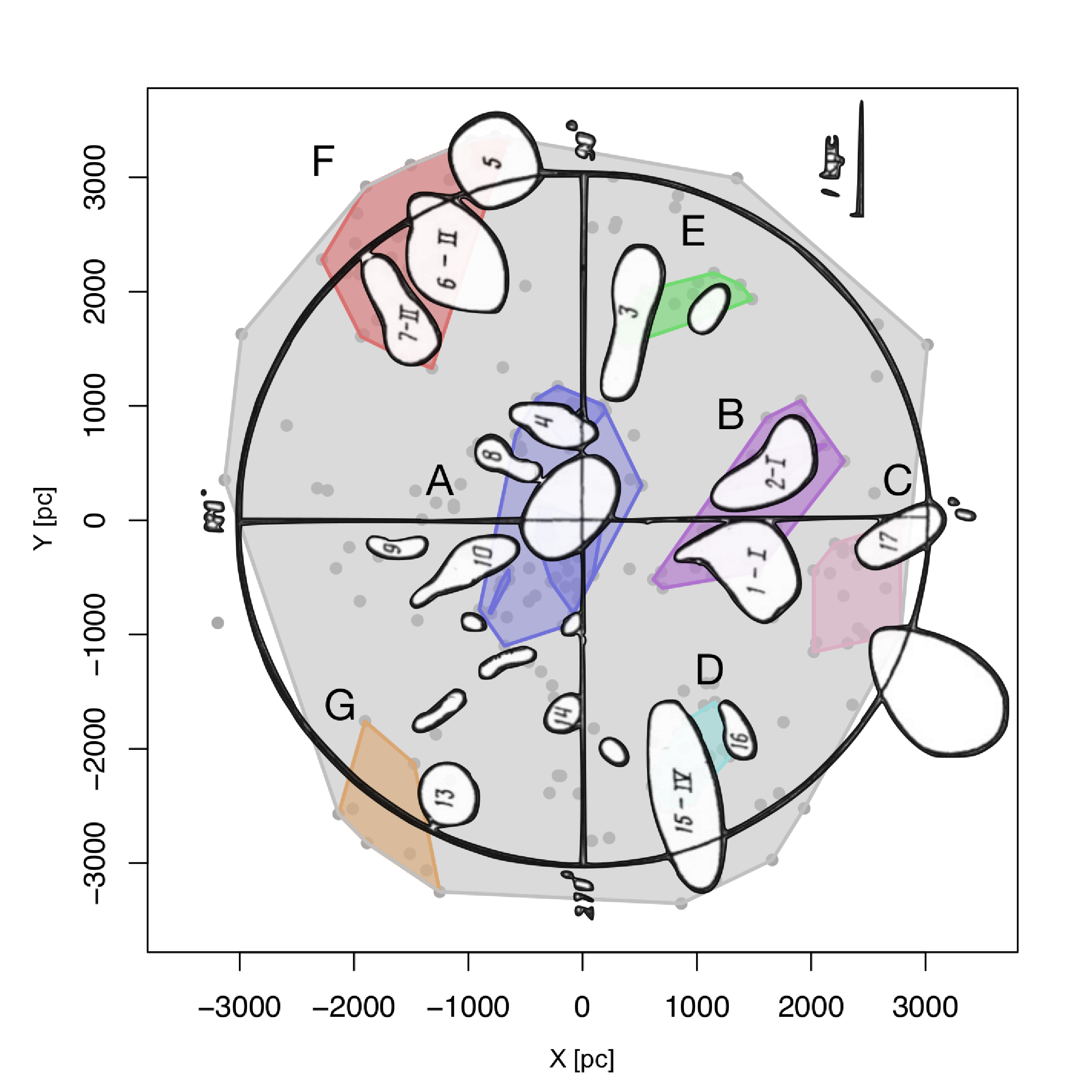}
\caption{Map of the structures retrieved in this work, as shown in the lower panel of Figure \ref{FigOpticsGral}. Fig. 2 from \citetalias{EfremovSitnik88} is overlaid for comparison, with their complexes and supercomplexes shown in white.}
\label{FigCompareEfremov}
\end{center}
\end{figure*}

To explore the G-YOC groupings distribution in more detail, we compare our results with those in the seminal work by \citetalias{EfremovSitnik88}. 
Figure \ref{FigCompareEfremov} shows the results of this comparison, where coloured polygons represent the convex hull of structures found in this work and the overlapping white blobs are the complexes and supercomplexes from \citetalias{EfremovSitnik88}. The results from \citetalias{EfremovSitnik88} show a striking resemblance to those obtained in this work, despite the obvious differences in the detection techniques, and nature and quality of data. In particular, most,  groupings found in this work can be linked to correspondent complexes from \citetalias{EfremovSitnik88}. All supercomplexes (marked with roman numerals in \citetalias{EfremovSitnik88}) correspond to structures of hierarchical level 1, with the exception of  structure D in the Sagittarius arm and which corresponds to the Carina~OB1 association. Structure E in the Cygnus area, corresponding to complex 3, also reaches hierarchical level 1.

The 2 kpc sideways box centered on the Sun shows the richest pattern, where the GB is included.
A Copernican education leads us to doubt singularities appearing at positions close to  observer's location. However, we will analyze in more detail the $ Z(X,Y) $ and $ V_Z(X,Y) $ maps in this region later.

Numerically, level 0 clusters ($ n = 9 $; the smallest scale of the hierarchy) have a size central value (median) of 630 pc, although within a wide range of diameters. Level 1 ($ n = 4 $) has a median of 1600 pc and the only level 2 cluster, located in the  solar vicinity, shows a 3D diagonal of 2300 pc. Level 0 and 1 clusters present similar sizes to those determined for Milky Way stellar complexes and supercomplexes in previous work, suggesting that these hierarchical levels  correspond to these structures. In particular, Efremov \& Sitnik \citepalias{EfremovSitnik88} estimated a range of between 150 and 700 pc for stellar complexes within a radius of 3 kpc around the Sun, which agrees with our measurements. They also provided a typical value of 1.5 kpc for supercomplexes, in perfect agreement with the central value for the level 1 clustering we consider associated with these structures. Structure A (Table \ref{tableOptics} and Figure \ref{FigOpticsGral}) with the largest size and highest degree of complexity deserves special mention. Its diameter of 2.3 kpc seems to relate it to those star-forming supercomplexes that are associated with sheared spiral arms \citep{ElmegreenEfremov1996}. These star-forming regions can reach sizes three times the width of the spiral arm and their origin could be associated with different gravitational resonances in the Galactic disk \citep{Elmegreens1992}. These supercomplexes also show a high degree of internal spatial hierarchy, as well as an intricate kinematic structure with the presence of several moving groups \citep{Lepine2018}. Structure A forms the core of the LA, the nature and origin of which is still a matter of debate. However, the size of A, its high degree of structural hierarchy, and the corrugations in $Z$ and $V_Z$, which show the largest amplitude and spatial scale, suggest an origin associated with gravitational traps \citep{Lepine2018, Tatiana2019}, plus a transient event (e.g., interaction with high velocity clouds, satellite galaxies, tidal tails, or dark-matter blobs) that would generate the local vertical and vertical-velocity higher amplitudes. 
The distribution on the plane of the clusters obtained with OPTICS indicates that the Perseus arm only contains the F supercomplex located in the II Galactic quadrant, between Galactic longitudes $ 100\degree \leq l \leq 140\degree $. Young clusters and massive stars appear to be more scarce  in the arm prolongation into the III Galactic quadrant \citep{Negueruela2003406, Cantat-Gaudin2020}. 
Clusters A and E seem to define the Orion-Cygnus (or LA) arm. Its natural continuation through the III Galactic quadrant seems to be the connection with cluster G, between the directions $ 240\degree \leq l \leq 250\degree $ as proposed by \citet{Moitinho2006} and later confirmed by \citet{Vazquez2008}. 
The Carina-Sagittarius arm is defined by structures B and D that are concentrations similar to those of massive stars and OB associations \citep{Wright2020, Pantaleoni-Gonzalez2021}. Concerning the spiral structure in the solar neighborhood, even if we only consider the seven larger structures in Figure \ref{FigOpticsGral} (bottom) it is difficult drawing a feasible scheme of such structure. Thus, perhaps we should think of a hierarchy of groupings, rather than of three spiral arms for analyzing the phase space of this Galactic region.   
The outstanding consistency between the global picture found in this work using state-of-the-art detection and analysis techniques with the results by \citetalias{EfremovSitnik88} highlights both the human potential for pattern detection, and the reliability of the parameter estimates  for clusters and associations, supporting their effectiveness to study in depth the large-scale structure of the Galaxy \citep[e.g.,][]{ElmegreenEfremov1997}.

\subsection{Kriging Results: Structure of the Galactic Disk}
\label{sub:krig_maps}

\begin{figure*}
     \centering     
    \includegraphics[width = 0.6\linewidth]{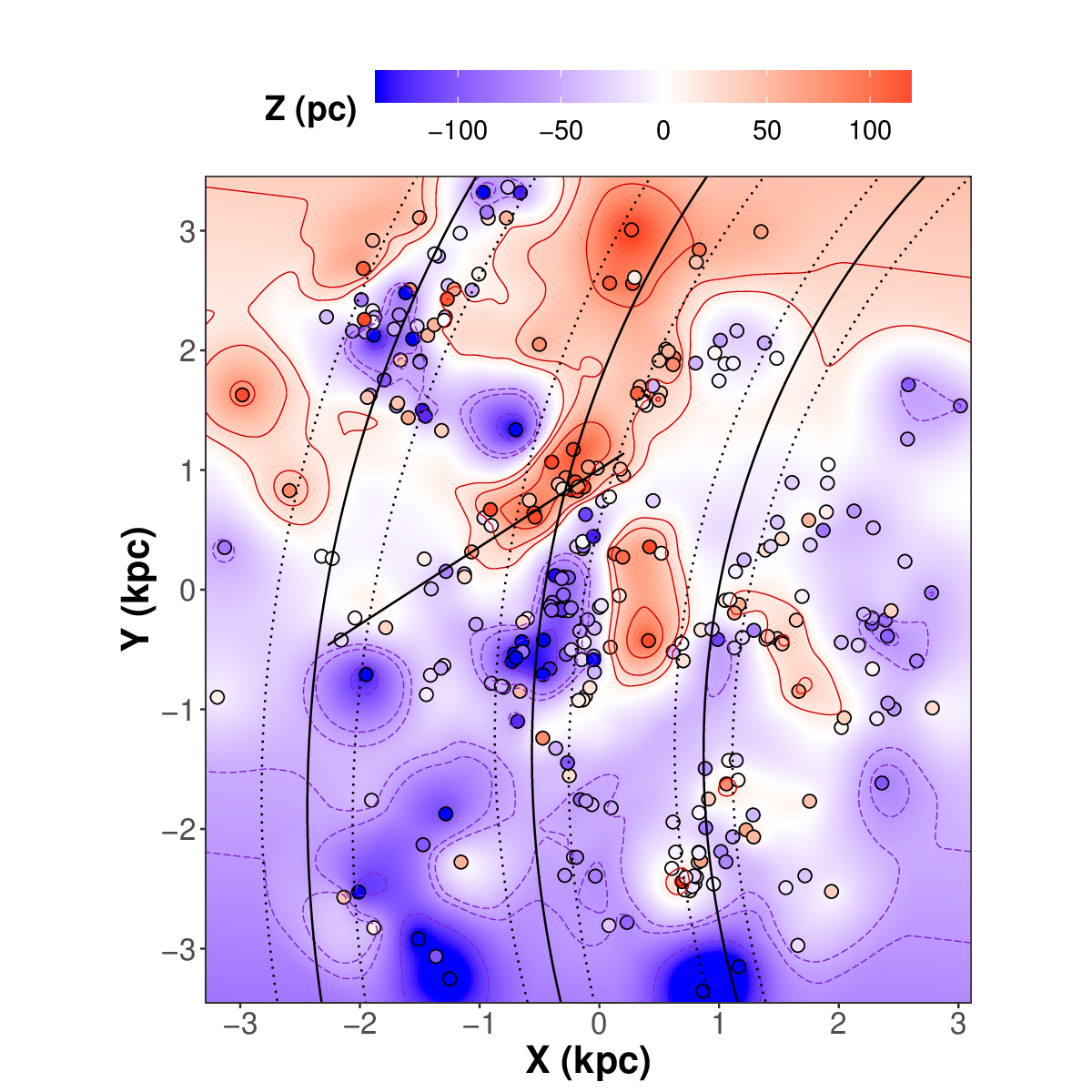}
    \includegraphics[width =0.6\linewidth]{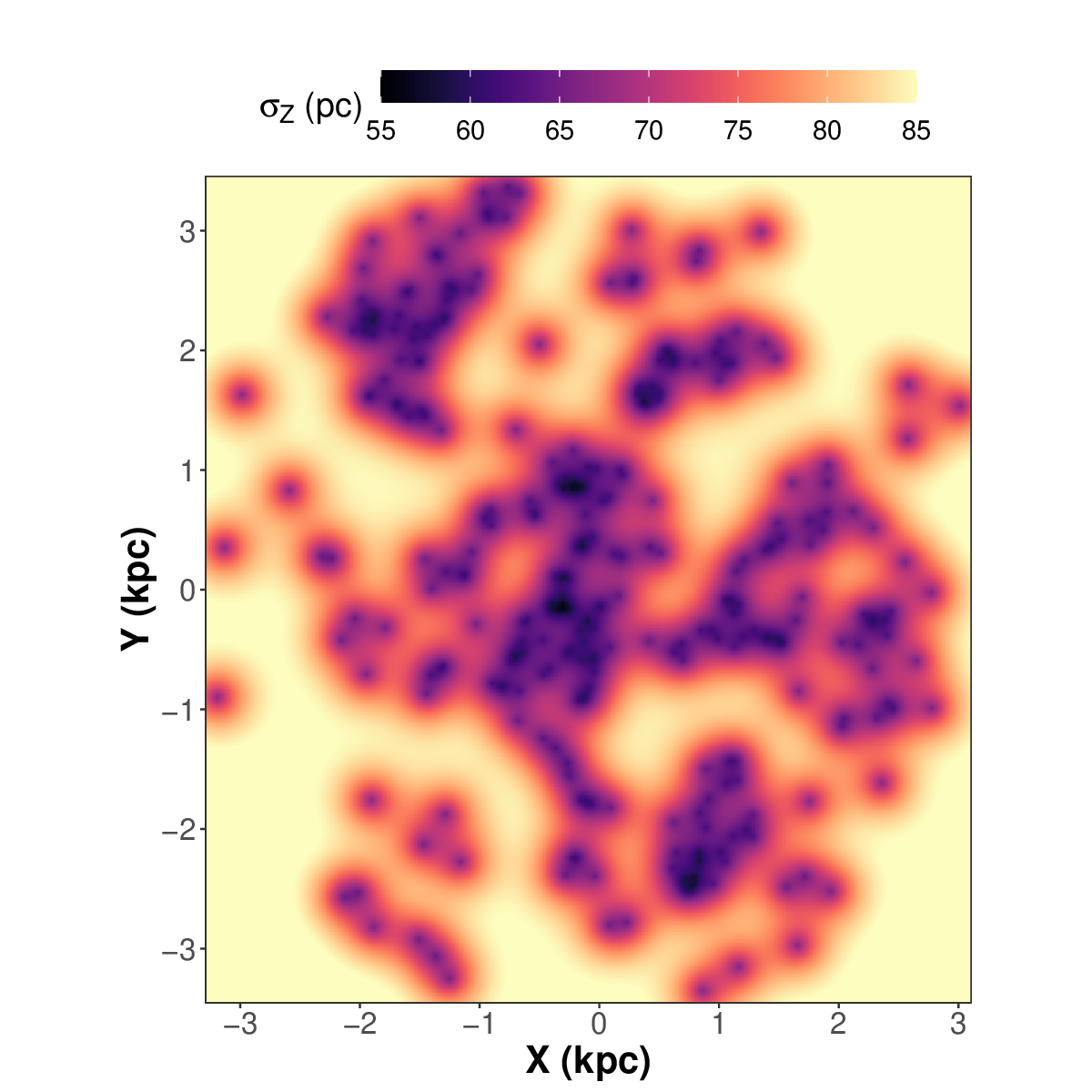}
     \caption{
     \textit{ 
    Top:} $ Z(X,Y) $ map or estimated $Z$ young Galactic disk structure, inferred  from the Kriging method. 
     Color dots represent the input data. The spiral arms are the same as represented in Figure \ref{fig:Young}, and the linear segment represents the Cepheus spur (see Table \ref{tab:spiral-arms-params}).
     \textit{Bottom:} The Kriging prediction standard deviation $\sigma_{Z}$ map (Equation \ref{eq:sigma_krig}), allowing a better assessment of the significance of the detected structures. 
     }
     \label{fig:KrigZ}
\end{figure*}

\begin{figure*}
     \centering     
    \includegraphics[width =0.6\linewidth]{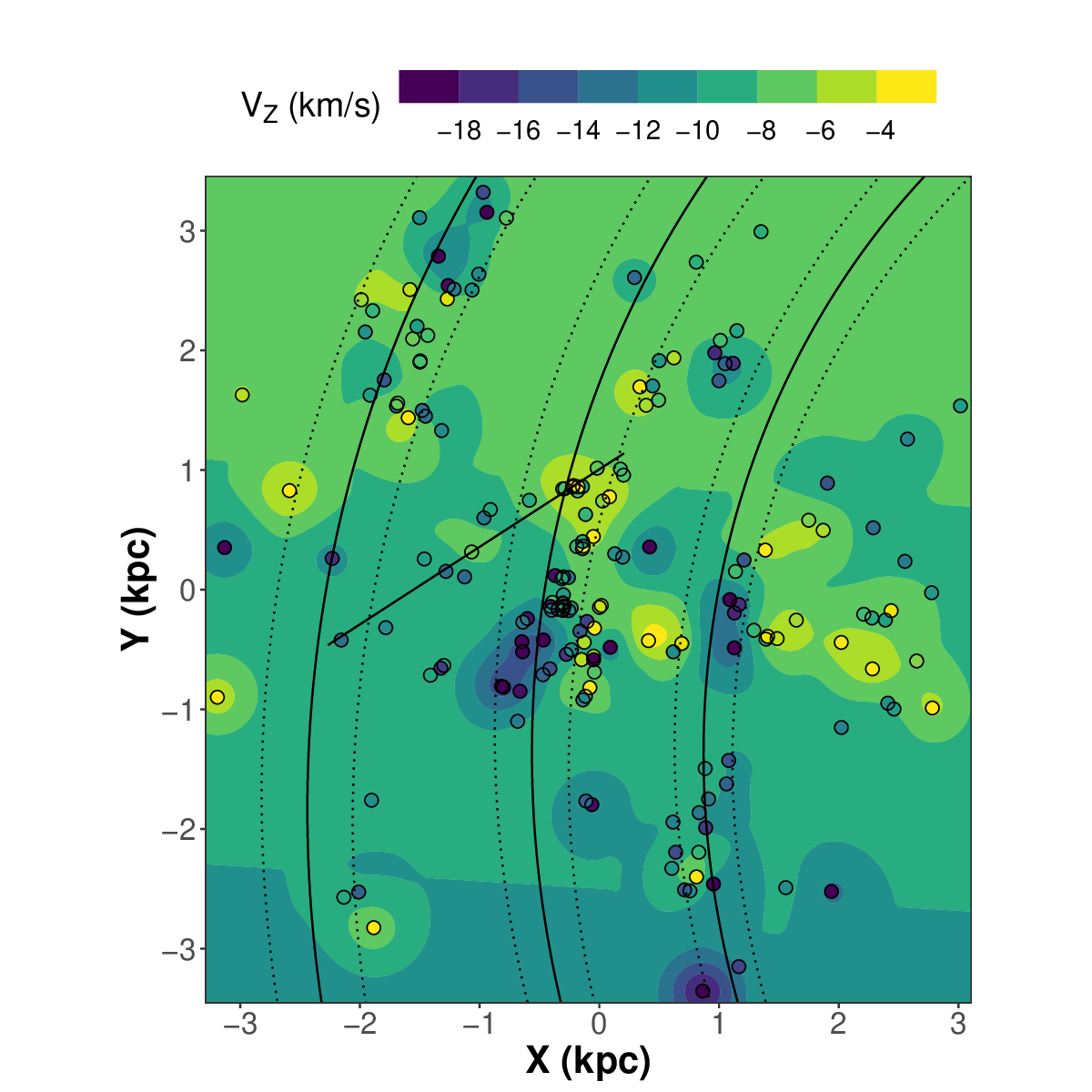}
    \includegraphics[width =0.6\linewidth]{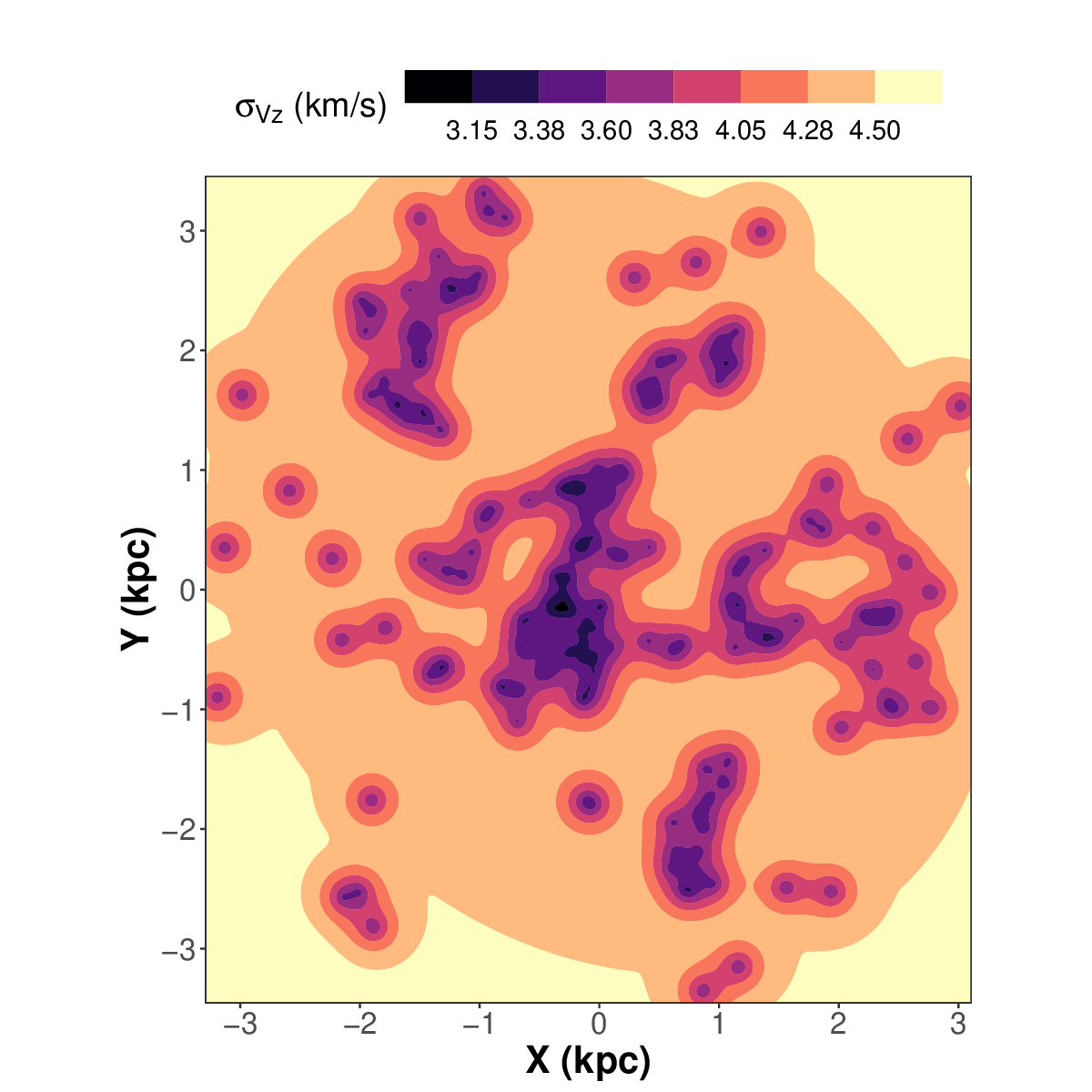}
    \caption{Analogously to Figure \ref{fig:KrigZ}, \textit{ 
    Top:} $V_Z(X,Y)$ map. For this case, we have taken a discrete scale since it suits better the minor deviation or range of this variable.
    \textit{
    Bottom:} The Kriging prediction standard deviation $\sigma_{V_Z}$ map (Equation \ref{eq:sigma_krig}).}
     \label{fig:KrigVz}
\end{figure*}

We finally show and describe the results obtained from the Kriging inference method presented in \S{\ref{subsec:Kriging}}. This interpolation approach provides the $ Z(X,Y)$ map and the $V_Z(X,\, Y)$ vertical velocity field of the Galactic disk defined by the G-YOC sample. As the YOCs are young objects that remain close to their birthplace and, in most cases, are associated with massive star-forming regions, these vertical structures can be related to the large-scale clustered star-forming activity in the Galactic disk. 
Figure \ref{fig:KrigZ} shows the $ Z(X,Y)$ Kriging spatial map, which provides an estimation of the 3D Cartesian structure of the disk drawn by the YOCs. Analogously, Figure \ref{fig:KrigVz} shows the corresponding one for $V_Z(X,Y)$. These maps correspond to a region within a 3.5 kpc radius around the Sun. We took a grid with steps at each ten pc to perform the Kriging estimation based on $Z$ and $V_Z$ measurements from the G-YOC, represented by filled circles in the latter figures. 
We also plot the position of the Cepheus spur and the spiral arms from  \cite{2021A&A...652A.162C} with fixed spiral widths based on \cite{2014ApJ...783..130R} to explore the phase-space behavior along with the spiral arms location, as it is discussed in next \S{\ref{sec:Discussion}}.  
Furthermore, the Kriging estimate at each grid point is mainly affected by the density of input data in its proximity. We can see this effect in the bottom plots of Figures \ref{fig:KrigZ} and \ref{fig:KrigVz}. 
The regions with lower variance correspond to the locations closer to the higher surface density of observed clusters\footnote{These Kriging matrices are available at CDS via anonymous ftp to \href{http://cdsarc.u-strasbg.fr}{cdsarc.u-strasbg.fr} (\href{ftp://130.79.128.5}{130.79.128.5}) or via \href{https://cdsarc.unistra.fr/viz-bin/cat/J/ApJ}{https://cdsarc.unistra.fr/viz-bin/cat/J/ApJ}.}. 

The $Z(X,Y)$ map shows a complex structure at different scales. A gradient of the vertical coordinate $Z$ is seen along the entire length of the $Y$ axis. Superimposed on this large-scale gradient there is a series of peaks and valleys, some of which show larger vertical displacements. The Galactic disk is corrugated with different amplitudes and spatial scales. One of the more striking substructures is located in the solar neighborhood, between $-1.0$ and $1.2$ kpc in the $Y$ coordinate, where the $Z$ vs. $Y$ gradient becomes deeper, reaching a $Z$ amplitude close to $200$ pc. The averaged $Z$ shows a negative value of $-20.9$ pc within a 50 pc radius around the Sun and of $-17.5$ pc on the whole region under study.  These heights are within the range of values obtained by other authors for different tracers with different ages and Galactic locations \citep[see][and references therein]{Elias2006a, MAAS2008,  Everall2022}.   
The $\sigma_Z$ map also provides a rough estimation of the Galactic plane height scale. This dispersion, as already pointed out in \citetalias{ACCD1991}, accounts for the intrinsic Galactic $Z$ dispersion, the precision of individual $Z$ measurements, and the inherent errors in the Kriging interpolation. 
We got an average estimate of $\sigma_Z = 77.4$ pc for the G-YOC sample, similar to that found by other authors for young stellar populations in this region \citep{MA2001, Piskunov2006, Zari2018, Kounkel2020}.

\citet{Pantaleoni-Gonzalez2021} state that the main features observed in the 3D map of their OB  stars sample are similar to those of the 3D map of YOCs derived 30 years earlier \citepalias{ACCD1991}. The same can be said of the comparison of the map obtained in this work and that of 1991:  a) there is a global $Z$ gradient along the axis parallel to Galactic rotation; b) the III Galactic quadrant appears to be located, in large extension, below the Galactic plane; c) the steeper spatial gradient in $Z$ appears close to the Sun; and d) the overall appearance of the disk is that of a corrugated surface, like an egg crate that has undergone huge deformation, giving rise to a not necessarily periodic distribution of peaks and valleys of different amplitude.

$V_Z(X,Y)$ shows some large- and intermediate-scale similarities with the spatial map but a clumpier kinematic structure at shorter scales. The velocity gradient observed in $V_Z$ is remarkable along the total length of the $Y-\,$axis. Overimposed to this general trend, a much steeper vertical velocity gradient is observed in the central region, limited by a 2.0 kpc side box. These steps gradients, at both  $Z$ and $V_Z$ variables, and their spatial coincidence with the location of the classical GB, are features that need to be inserted into any global scenario for explaining the phase structure of the young Galactic disk. 

\begin{figure*}
    \centering
    \subfigure[$Z$ and $V_Z$ profiles along the Perseus, Local, and Sagittarius spiral arms, from left to right.]
    {
        \includegraphics[width=1.00\linewidth]{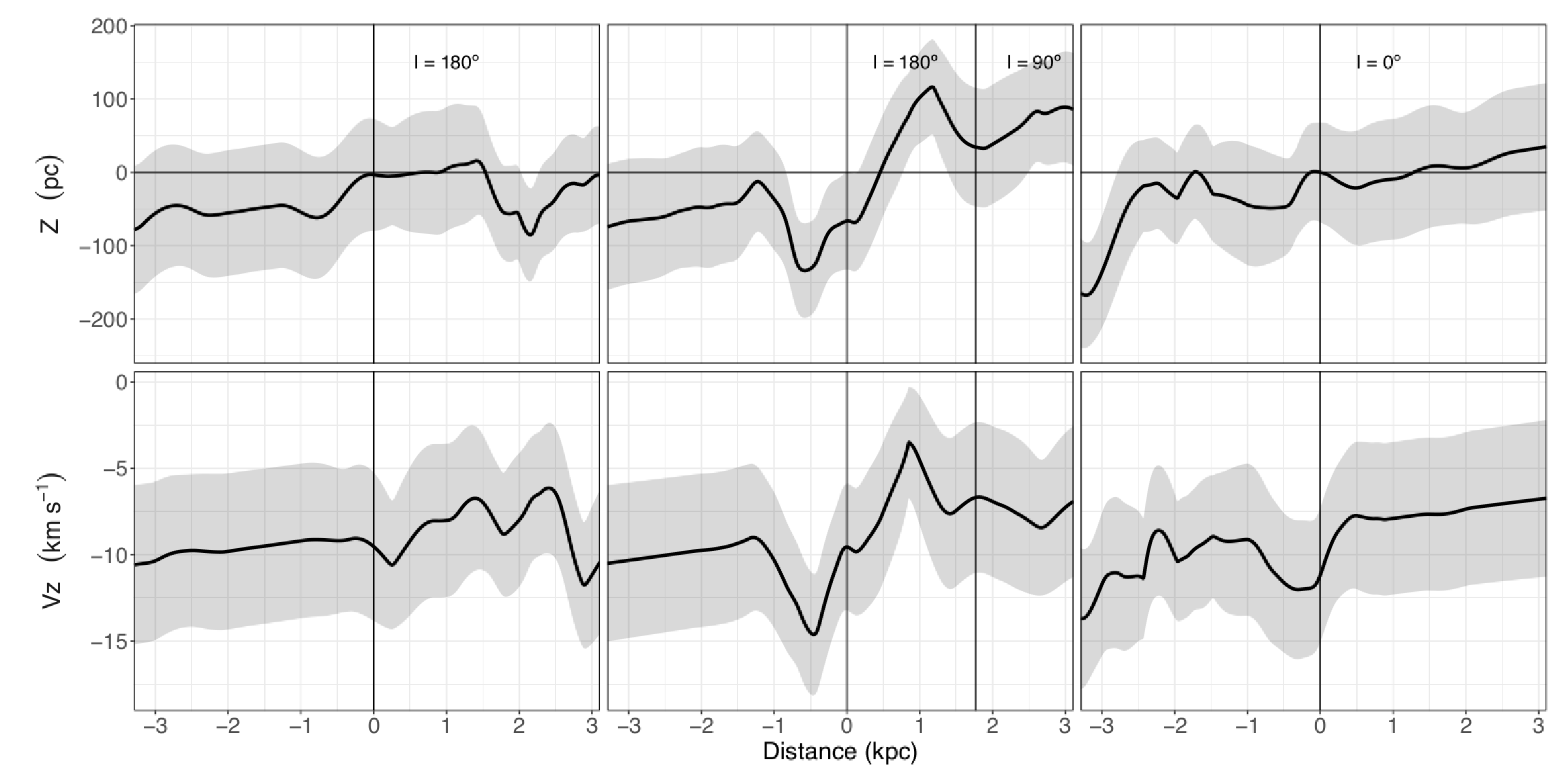}
        \label{fig:linearfeatures1}    
    }
    \\
    \subfigure[Profiles along the Cepheus spur.]
    {
        \includegraphics[width=0.28\linewidth]{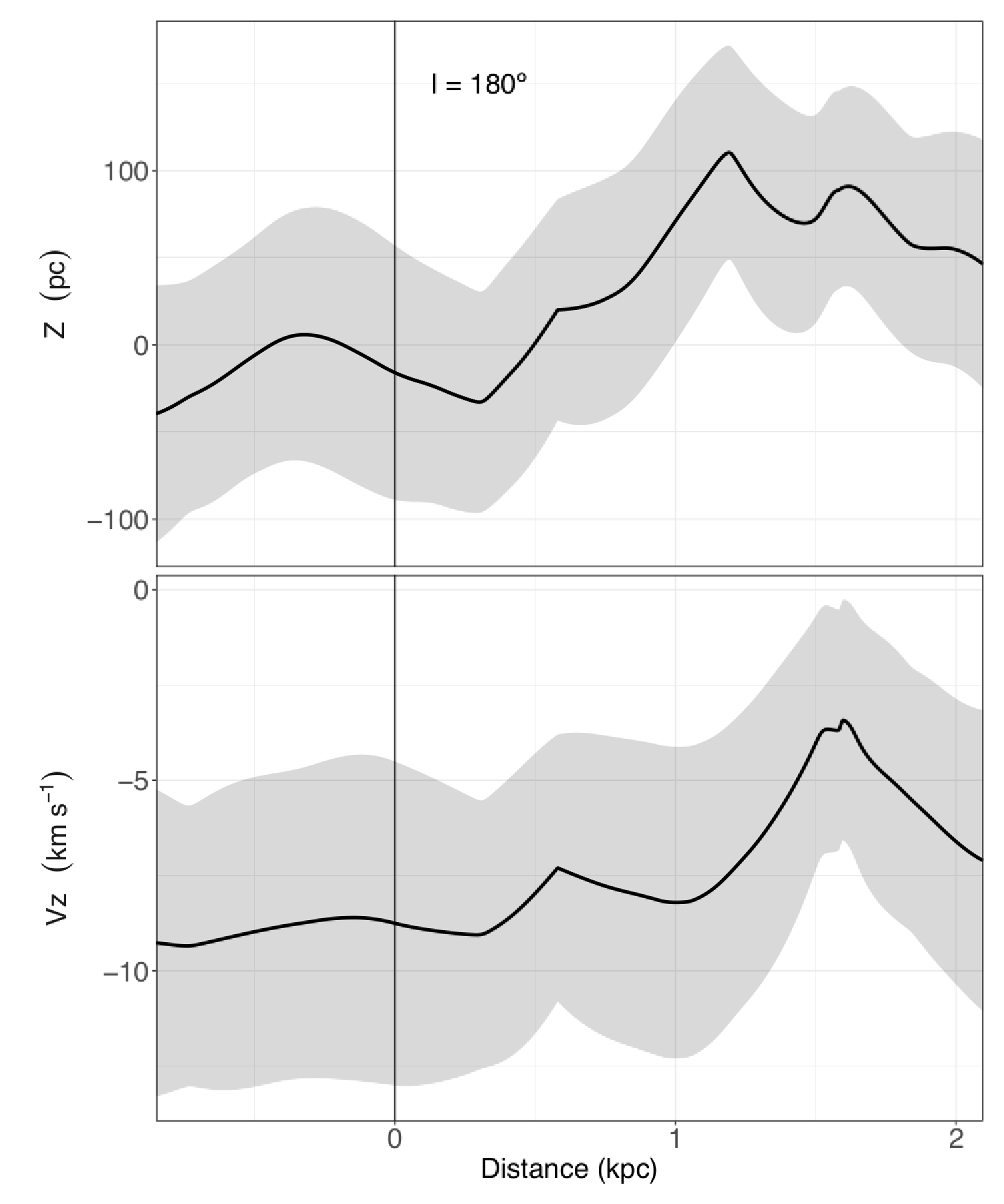}
        \label{fig:linearfeatures2}
    }
    \qquad
    \subfigure[Profiles along the Main diagonal.]
    {
        \includegraphics[width=0.65\linewidth]{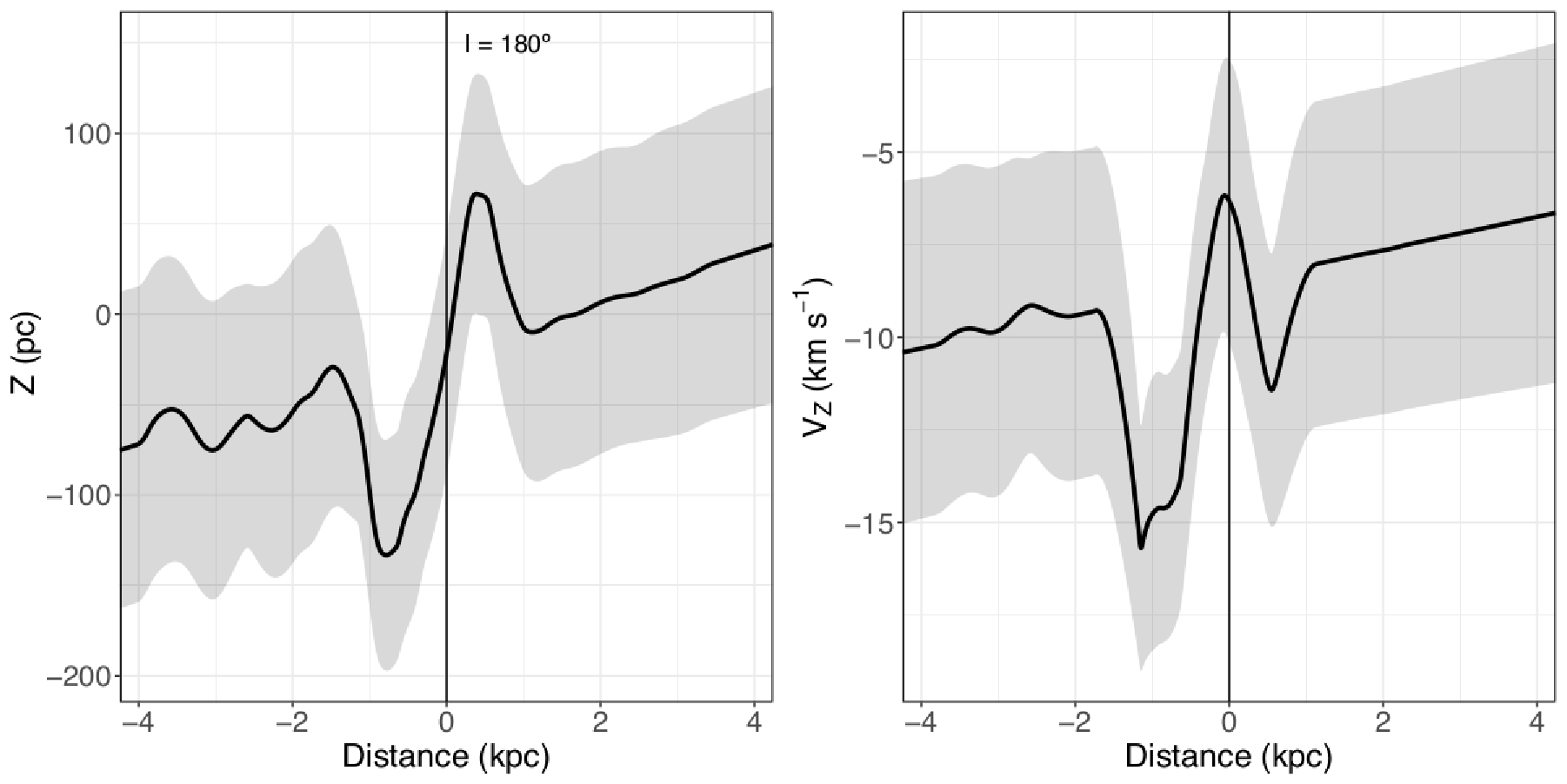} 
        \label{fig:linearfeatures3}
    }
    \caption{$Z$ and $V_Z$ profiles along different spatial structures or features. The shaded area accounts for the corresponding dispersion $\sigma_Z$ and $\sigma_{V_Z}$. The distances are measured along the cut, with the origin located at $Y=0$ in heliocentric spatial coordinates.}
    \label{fig:linearfeature}
\end{figure*}

Finally, we show some cuts of these maps along some selected directions on the Galactic plane. Figure \ref{fig:linearfeature} shows the profiles of $Z$ and $V_Z$ throughout the Perseus, Local, and Sagittarius spiral arms, the Cepheus spur, and the Main diagonal $ (Y=X) $. 

The top panels of Figure \ref{fig:linearfeatures1} display the corresponding $Z$ profile, as thick black lines, along the three spiral arms. The shaded area accounts for the associated Kriging dispersion $\sigma_Z$. In contrast to Perseus and Sagittarius arms, LA shows a sharper $Z$ gradient in the central region. 
The expected corrugations observed in other works \citep[e.g.,][]{SF1986, alfaro1992} are also visible in these profiles.

For the vertical velocity, we observe that the global trend is a $V_Z$ gradient along the arms, being less pronounced in the case of the Perseus arm. The $V_Z$ profiles show their maximum slope in the regions close to the Sun, as with the $Z$ spatial coordinate. Other special features, as Cepheus spur (Figure \ref{fig:linearfeatures2}), and the Main diagonal ($Y = X$; Figure \ref{fig:linearfeatures3}), also host a similar behavior in $Z$ and $V_Z$, with undulations in both variables.  These results are discussed in more detail in the following section {\ref{sec:Discussion}}. 


\section{Discussion}
\label{sec:Discussion}

Both the 3D spatial distribution and the vertical-velocity field have three common features:
\begin{itemize}

    \item A large-scale spatial variation spanning the entire area under study, which has been historically associated with the Galactic warp \citep[][among others]{Levine2006, Romero2019, Cheng2020}.
    
    \item A central region, 2.0 kpc on a side, with a high degree of 2D, 3D, and $V_Z$ structure, which includes the traditional GB \citep{Elias2006a, Comeron1999, Romero2019} and dominates the LA vertical profile \citep{dixon1967}. 
    
    \item  Superimposed on the large-scale spatial variation, smaller amplitude ripples are present, for both $Z$ and $V_Z$, in any direction we observe. These corrugations affect the entire disk and cannot be considered specific to particular directions, although some gas and young stellar objects alignments, such as Cepheus spur \citep{Pantaleoni-Gonzalez2021}, merit a more detailed analysis.
\end{itemize}

The profiles of $ Z(X,Y) $ and $ V_Z(X,Y) $ along the Main diagonal $ Y = X $ clearly show the three features highlighted above. In Figure \ref{fig:linearfeatures3}, a well-defined gradient with the distance in $Z$ is observed along the Main diagonal, which can be estimated as $ 14.78 \pm 1.56 $ \pckpc for the whole linear segment. 
The same pattern is observed in the variation of $V_Z$ from the I to the III Galactic quadrant, with a global slope  of $ 0.50 \pm 0.36 $ \kmskpc.
The central region shows a steeper slope in $Z$ and $V_Z$ ($ 170 \pm 9.7 $ \pckpc and $0.92 \pm 0.26 $ \kmskpc, respectively), within a radius of 1 kpc around the Sun. 

Finally, the existence of $Z$ and $V_Z$ ripples with amplitudes of  $\sim50$ pc and $\sim3 $ \kms, respectively, and spatial scales between 1 and 3 kpc, dominate the whole disk. We are using the $Z$ and $V_Z$ Main-diagonal profiles to illustrate the three different phenomenologies observed in this Galactic region with their different scales and amplitudes.

\subsection{Vertical Spatial and Velocity Large-scale Variations: Can the Warp be Detected in this Galactic Region?}
\label{subsec:warp}

\cite{Romero2019} already reported asymmetry in the median $Z$ about the solar-galactocentric line ($\phi = 180\degree$, anticlockwise), for both OB (age $<$ 1 Ga) and older RGB stars, being the warp-down amplitude (at $\phi \geq 180 \degree$) larger than the warp-up (at $\phi \leq 180 \degree$).  This asymmetry can be interpreted as a signature of the Galactic warp lopsidedness.   
They also found an age dependence with the amplitude of the stellar warp, increasing with the age of the tracer population. 
In the case of the warp kinematics, the median proper motion in Galactic latitude also shows a stripe in the warp-down amplitude. 
Our maps are limited to the central 3.5 kpc around the Sun. This fact, together with the age differences of the samples, could yield different views. 
However, we see a good agreement with our kinematic results. Overall, if we focus on the clear gradient of $V_Z$ and $\mu_b$ along the $Y$ axis, restricting the analysis to the same location (see their Fig. 8, left panel).
The $ Z(X,Y) $ and $ V_Z(X,Y) $ maps show characteristics compatible with the detection of a warp defined by YOCs according to the features presumed to be associated with stellar warp in previous works:

\begin{enumerate}
    \item Galactic north-south asymmetry in the vertical velocity $V_Z$ distribution.
    \item Asymmetry in tracer density and $Z$ between I and II and  III and IV Galactic quadrants, being more evident between II and III quadrants.
    \item Undulating behavior of $Z$ and $V_Z$ with the Galactocentric radius $R$.
    \item Vertical velocity gradient $V_Z$ with the spatial component $Y$.
    \item Warp amplitude dependence on the age of the tracers, decreasing for the younger ones.
\end{enumerate}

This list includes and summarizes a wide variety of works carried out with different observational samples, which, in many cases, implies different age ranges and different Galactic volumes.  
Now, all these observational features can be seen, for the first time, in the G-YOC distribution described by the $ Z(X,Y) $ and $ V_Z(X,Y) $ maps.

\begin{figure}[t]
\centering
\includegraphics[width = 0.45\textwidth]{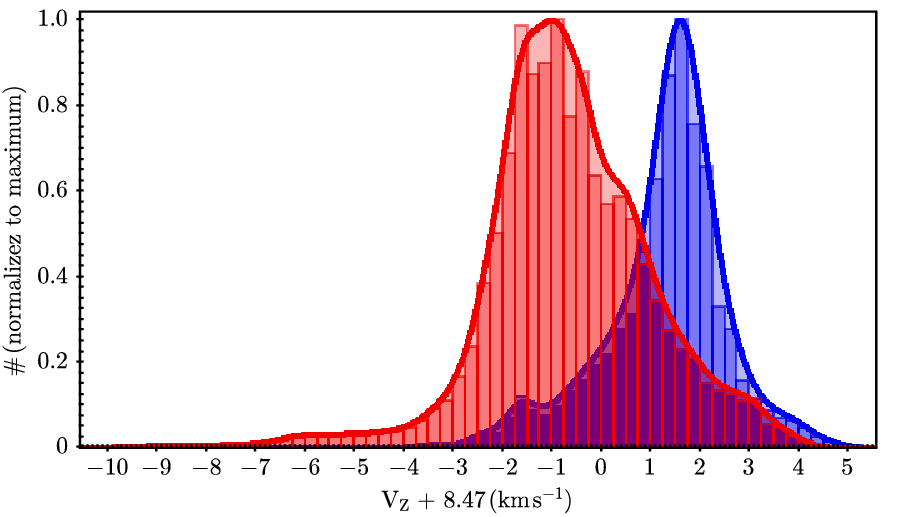}
\caption{Histogram of $V_Z + 8.47$ \kms. This constant is the median of $V_Z$ in the area under study. We have shifted $V_Z$ for a better visualization of the vertical velocity behavior in both hemispheres.  In red $V_Z$ for $Z\ \leq$ 0 and blue for $Z\ >$ 0.}
\label{fig:Histogram}
\end{figure}

When analyzing the $ Z(X,Y) $ and $ V_Z(X,Y) $ maps to study their relationship with the Galactic warp we must take into account two fundamental variables: the age of the tracers and the area under study. The latter puts constraints on the range of Galactic radii analyzed. 
Recent studies of the Galactic warp  \citep[e.g.,][]{Romero2019, Cheng2020}, already based on \textit{Gaia} data releases, concluded that both $ Z(X,Y) $ and $ V_Z(X,Y) $ maps present a high degree of substructure that cannot be solely explained by classical Galactic warping models.

These previous results highlights the importance of accurately knowing the samples' properties and comparing the results obtained in previous work based on tracers of a similar nature. 

In Figure \ref{fig:Histogram} we show the histogram of the vertical component of the velocity, $V_Z$, for the two Galactic hemispheres ($ b > 0 $; North, and $b \leq 0 $; South) within the  area under study. We have added a constant of $8.47$ km (median $V_Z$) to the $V_Z$ values to  better visualize that the YOCs vertical kinematics of the northern Galactic hemisphere  is dominated by positive values. In contrast,  those in the southern hemisphere show a negative distribution mode.

\begin{figure}
      \includegraphics[width=0.45\textwidth]{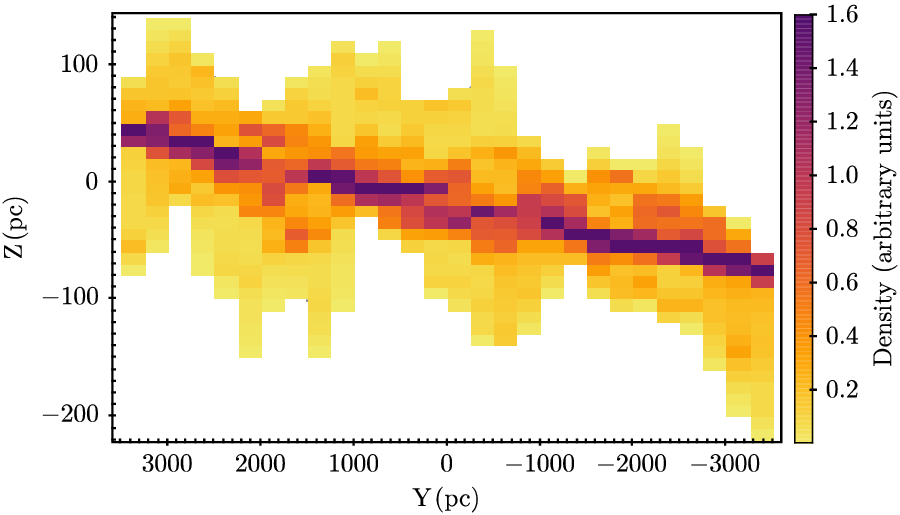}
      \includegraphics[width=0.45\textwidth]{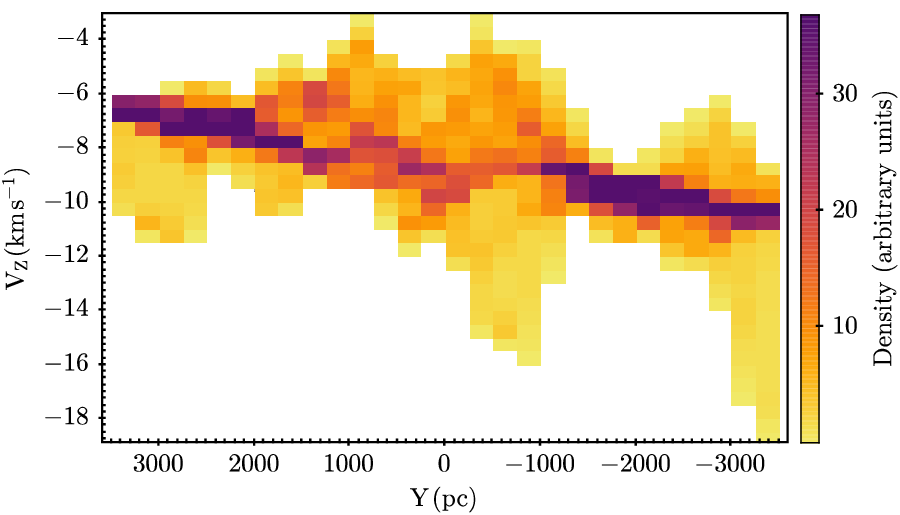}
    \caption{On the top, $ Z(X,Y) $ map projected onto $ Y $ axis, both in pc units. Negative $Y$ values correspond to III and IV Galactic quadrants. A gradient $ Z(Y) $ is clearly visible, overimposed by  a wave-like structure.
    On the bottom, $V_Z(X,Y)$ map in \kms  projected onto $Y$ axis in pc. Negative $Y$ values correspond to III and IV Galactic quadrants. A  gradient $V_Z(Y)$ is the main feature of the plot.}
    \label{fig:warp_Z_Vz_Y}
\end{figure}

Figure \ref{fig:warp_Z_Vz_Y} shows the $ Z(X,Y) $ and $ V_Z(X,Y) $ maps projected onto the $ Y $ axis. The projections of the Kriging maps show again a similar scheme to the one observed in the Main diagonal: a well-defined large-scale gradient with a superposition of peaks and valleys of different amplitudes, densities, and spatial scales, finding the steepest slope in locations close to the Sun.

\begin{figure*}
\centering
\includegraphics[width = 0.49\textwidth]{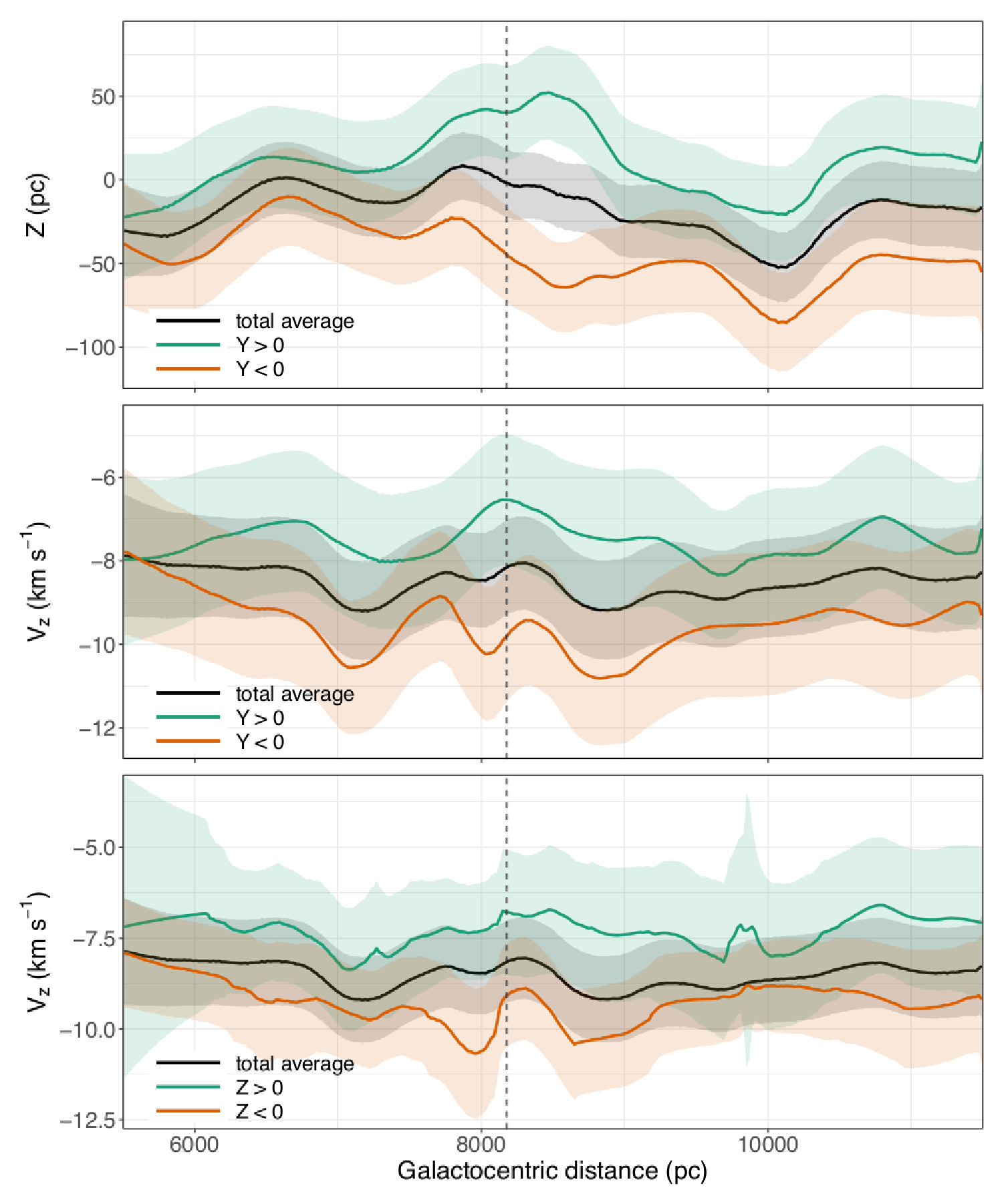}
\includegraphics[width = 0.49\textwidth]{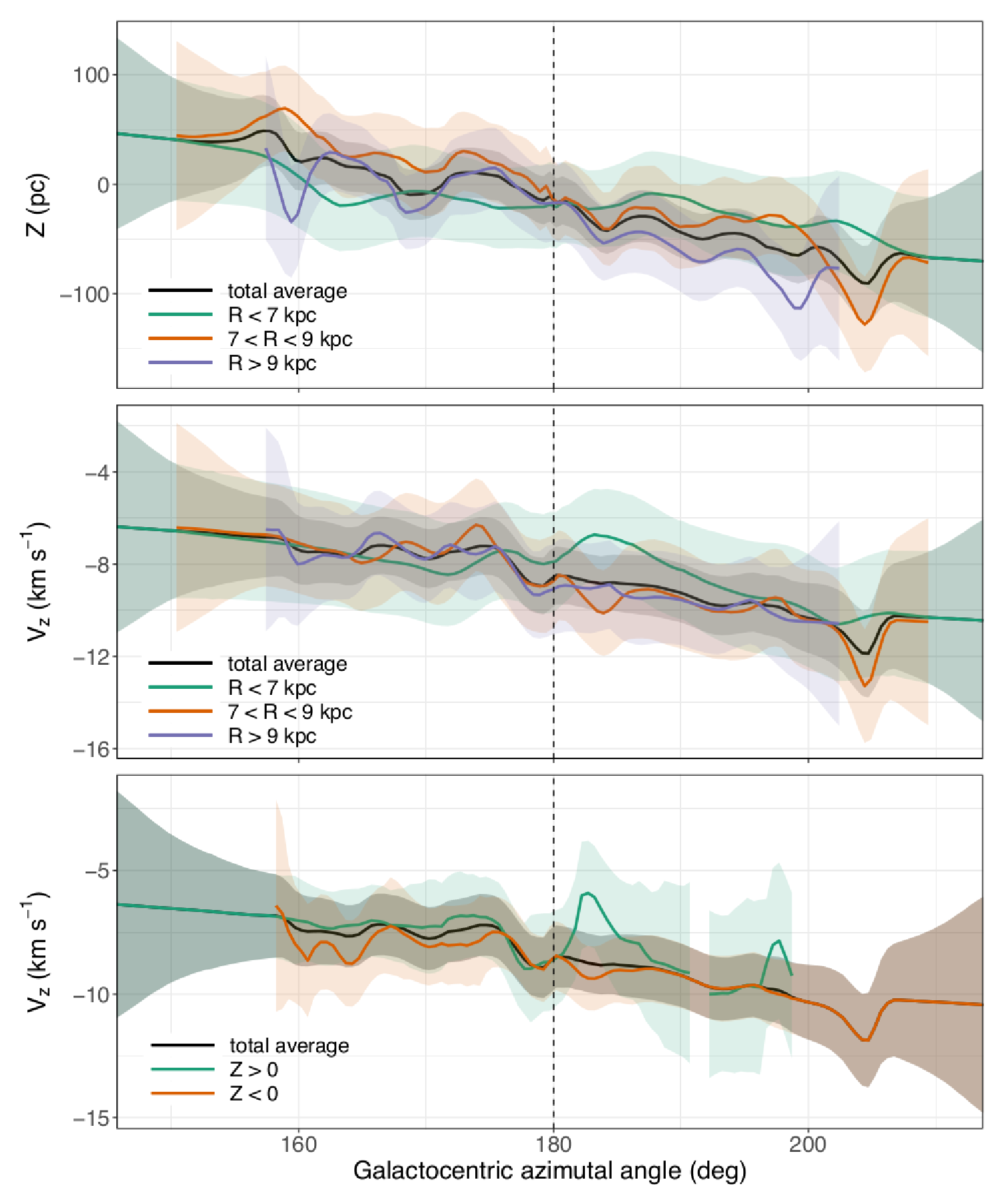}
\caption{Weighted average Kriging estimations for $Z$ (pc) and $V_Z$ (\kms) as a function of the Galactocentric radius, ($R$; bins of 20 pc) and the Galactocentric azimuthal angle ($\Phi$; bins of 0.5$^{0}$) respectively. 
$\sigma^{-1}_i$ is taken as weight.  
The shaded areas correspond with the standard deviation of the weighted average, assuming correlated measurements. 
The vertical dashed line accounts for the Sun location.}    
\label{fig:warp}
\end{figure*}

The same fact is observed in Figure \ref{fig:warp}, where the moving average of $V_Z$ vs. $R$ has been plotted for the northern and southern hemispheres, as well as for the entire map. It is clearly observed the separation  between $Z$ y $V_Z$  for the I and II Galactic quadrants and the III and IV ones. 
Figure \ref{fig:warp} shows the weighted average of $ Z(R) $ and $ V_Z(R) $ profiles on the left, and $ Z(\Phi) $, $ V_Z(\Phi) $ on the right. We took bins of 20 pc and bins of $0.5$ degrees for $R$ and $\Phi$, respectively, to compute the weighted average of the $Z$ and $V_Z$ maps (Figures \ref{fig:KrigZ} and \ref{fig:KrigVz}). The overall weighted mean, encompassing the whole region, is plotted for all cases with a black line together with the weighted mean for different cases for each variable. 
Weights are the normalized inverse of the Kriging standard deviation $w_i = \frac{\sigma_i^{-1}}{\sum_{i=1}^n \sigma_i^{-1}}$, and the global uncertainty or variance of the weighted average is computed as $\sigma^2_{avg} = w \Sigma w^T $, where the covariance matrix $\Sigma$ is estimated by the sample cross covariogram provided by \texttt{Gstat}. At larger $R$, uncertainty increases due to the scarcity of observations at those Galactocentric radii.  

If we look at the right plot of Figure \ref{fig:warp}, where the variations of the vertical coordinate with Galactocentric azimuth ($\Phi$) for different intervals of $R$ are shown, we can observe that there is a gradient of $Z$ and $V_Z$ with $\Phi$ that becomes more marked as $R$ increases. These results are in general agreement with those of previous studies \citep{Levine2006, Romero2019,  Wang2019, Cheng2020}, but with a caveat, here they have been obtained for a sample of young stellar clusters with ages less than 50 Ma, which is a useful observational constraint to narrow down the possible origins of these structures.

Previous studies have pointed out that the Galactic warp, defined by the young stellar population, seems to show its southern larger deviations between Galactic longitudes $230\degree$ and $260\degree$ \citep{Moitinho2006, Vazquez2008, Carballo2021}.

Answering the question that titled the subsection: Yes, the spatial and kinematic distributions of the YOCs show average values along different variables that can be associated with warping. However, the map drawn by $Z$ and $V_Z$, if we look at short spatial scales, is very far from resembling what the models predict \citep[see][for different mock models ]{Romero2019}, both in shape and magnitude.  

\subsection{The Solar Neighborhood}

The central region of the area under study requires a more detailed analysis. We define this region as a 2$\times$2 kpc square centered on the Sun. The square box contains the highest amplitudes in $Z$ and $V_Z$ (Figure \ref{fig:linearfeatures3}) found in our study, which already singles it out as a peculiar region. Regarding the distribution of YOCs analyzed in \S  \ref{subsec:clustering}, this box is preferentially associated with the supercomplex A, which presents a 3D diameter of 2.3 kpc. The supercomplex A shows the highest spatial hierarchy (level 2), except for the one corresponding to the total area (level 3), and also presents the highest density of YOCs, especially in the Orion region \citep[as already noted by other authors:][]{Piskunov2006, Bally2008, EACC2009}.

Within the complicated phenomenology of phase space in the solar neighborhood, the central box includes the LA most populated segment, and the classical GB. Figures \ref{fig:KrigZ} and \ref{fig:KrigVz} show the $ Z(X,Y) $ and $ V_Z(X,Y) $ maps, respectively, for this region, where we clearly appreciate the presence of, at least, three substructures in $ Z(X,Y) $ whose relative position on the plane explains and defines both, the  $Z$ undulation along the LA segment \citep[e.g.,][among many others]{dixon1967, quiroga1974, SF1986, Alves2020, Pantaleoni-Gonzalez2021}, and the classical inclination of the GB with a line of nodes at $l\ \approx\ 280\degree$ \citep{Comeron1994b, Elias2006a}.

The vertical velocity field, $V_Z$, also presents a similar pattern with four different clumps, which draws a clear variation of $V_Z$ along the LA central segment  (see Figure \ref{fig:linearfeatures1}  for a longitudinal cut along the arm). In turn, we observe a $V_Z$ gradient between the two inter-arm substructures, which had been previously associated to an oscillation of the GB around the solar Galactocentric  axis \citep{Comeron1999}.

Without entering into other considerations about the nature and origin of the spiral arms, it is evident that the amplitudes of the spatial Cartesian coordinate, $Z$, and of the vertical velocity component, $V_Z$, as well as the high concentration of YOCs in the region, suggest that large amounts of energy and momentum have come into play in a relatively small spatial volume. 

The physical mechanisms proposed in the literature for injecting these enormous amounts of momentum and energy into the solar neighborhood have changed over time \citep[see e.g.,][for a review of possible corrugation's generators at that time] {AlfaroEfremov1996}. New sources of momentum and energy have been incorporated in recent years, which together with the classical ones could be summarized as: supernova explosions \citep{Olano1982}, collisions with high velocity clouds \citep{Franco1988, Comeron1994a}, encounters with dark-matter blobs \citep{Bekki2009}, and interactions with tidal streams \citep[][and references therein]{Laporte2018, Ruiz-Lara2020, Tepper2022}, among others. This wide phenomenology can inject, into the interstellar medium, energies higher than 10$^{51}$ erg in a single event. Note, for example,  the case of NGC~6946 with HVCs' kinetic energies over 10$^{53}$ erg  \citep{Kamphuis_tesis1993}, able of displacing giant molecular clouds at heights of 150 pc from the formal Galactic plane. The existence of empty bubbles in the solar neighborhood, bordered by gas clouds and young stellar groups, and generated by any of these mechanisms has been detected, and described in several papers  \citep [see e.g.,][and references therein]{Lallement2003, Gontcharov2019, Zucker2022}. The subsequent dynamical evolution of the stellar population would explain the formation of the observed tilted  plane associated to GB.

Otherwise, there is kinematic information that shows the solar neighborhood as a collection of moving groups rather than as a single structure (disk or ring) with a coherent motion able of surviving longer than a few Ma \citep {Elias2006b, EACC2009, Lepine2018}. Two OB associations (or better, two groups of star-forming regions: Orion and Scorpius-Centaurus) located at the extremes of the $Z$ interval in the local spatial distribution of OB stars, show very different properties. While Orion has a high concentration of star clusters, Sco-Cen shows a more dispersed star formation, where it is difficult to find any substantial cluster in the optical range \citep {EACC2009}. Likewise, the kinematics of Sco-Cen cannot be explained only as an expanding star-forming region subjected to a stationary Galactic gravitational field \citep{MAF1999, Antoja2008, EACC2009, Lepine2018}. Sco-Cen OB associations appear to be associated to the Pleiades moving group, which includes the star cluster of the same name \citep{Chen1997, Asiain1999, Montes2001, Makarov2007}. The cluster and the associations are spatially separated by more than 300 pc and show ages differing by a factor of 10. The formation of these moving groups in the solar neighborhood, with an age range exceeding, in some cases, several hundred Ma, can be explained by dynamical resonances, generating gravitational traps where several moving groups can orbit for some Ga in a bounded region of the solar neighborhood \citep{Fux2001, Antoja2008, Minchev2010, Lepine2018, Tatiana2018, Barros2020}.

On the other hand, the main phase-space feature in this region is a spiral-shaped structure in the $Z-V_Z$ plane, whose contours are even sharper when modulated by the in-plane velocity components \citep{2018Natur.561..360A, Tian2018, Tatiana2019, Bland_Galah2019, Li2021}. The stellar samples that draw these spirals are mainly formed by stars with parallax errors lower than 20\% (avoiding negative parallaxes) and with radial velocities  included in \textit{Gaia}~DR2 and/or in any other of the extensive radial velocity surveys listed in \S{ \ref{subsec:input-data}}. Although there are  differences in the selection criteria of the distinct  samples, they show some common properties \citep[e.g., see][for details]{Tatiana2019}. The samples are representative of a stellar population with a large proportion of late-type stars over a very wide age interval, spanning the age of the Galactic disk. The wide range of ages drawing this structure would suggest that the initial phase-space pattern (if any) is not completely erased in a time period  shorter than the typical age of the sample, or, on the contrary, that the Galactic driver needs this long time to generate the peculiar \textit{snail  shells}  in the vertical phase space.

Other works have demonstrated that the formation of vertical phase spirals is independent of the Galactocentric radius \citep[][and references therein]{Li2021} and of the chemical properties of the sample \citep{Bland_Galah2019}. In other words, the  mechanism generating these structures seems to be acting over a long time period ($\sim$\ disk's age), and over the entire Galactic disk.

\begin{figure}
\includegraphics[width = 0.49\textwidth]{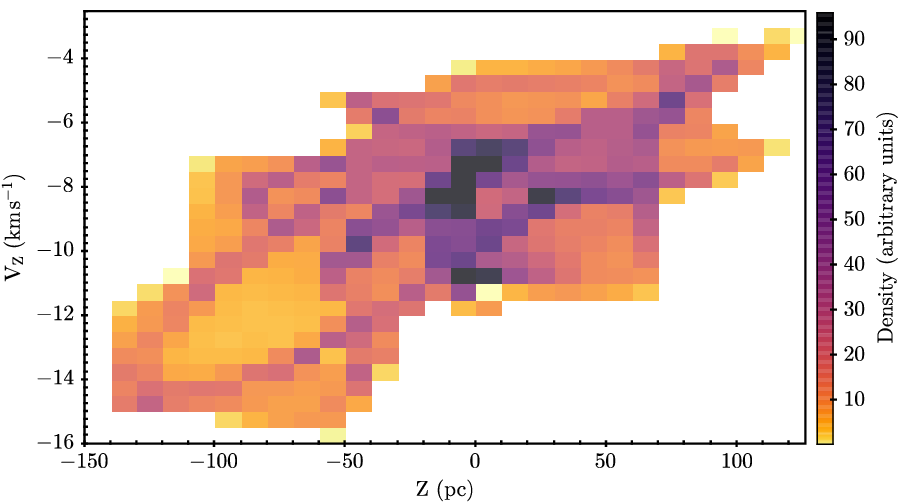}
\caption{$ V_Z $ vs. $ Z $ for the central box, 2.0 kpc on a side, containing stellar supercomplex A. This area includes the most populated LA segment, as well as the classical GB. Color code indicates map-points density in arbitrary units. A clear correlation between both phase-space coordinates is observed.}
\label{fig:central_box_phase}
\end{figure}

The $ Z(X,Y) $ and $ V_Z(X,Y) $ fields defined by the YOCs show a clear correlation between $V_Z$ and $Z$ for the central box in the solar neighborhood (Figure \ref{fig:central_box_phase}). If we compare this plot with the spirals observed in much older stellar samples \citep{2018Natur.561..360A, Tian2018, Tatiana2019, Bland_GALAH2019b}, we see that the linear structure, $ Z $ vs. $ V_Z $, depicted by the YOCs appears to be within the central region of the $snails$ detected in these works. 

\citet{Tian2018} have used \textit{Gaia}-DR2 astrometric data, together with LAMOST radial velocities, to analyze the vertical phase in different age groups. Their youngest sample (age $ < 0.5 $ Ga) is ten times older than G-YOC, and the corresponding vertical phase does not seem to show any well-defined structure, contrarily  to the roughly  linear relationship obtained in this work.

\begin{figure}
\includegraphics[width = 0.49\textwidth]{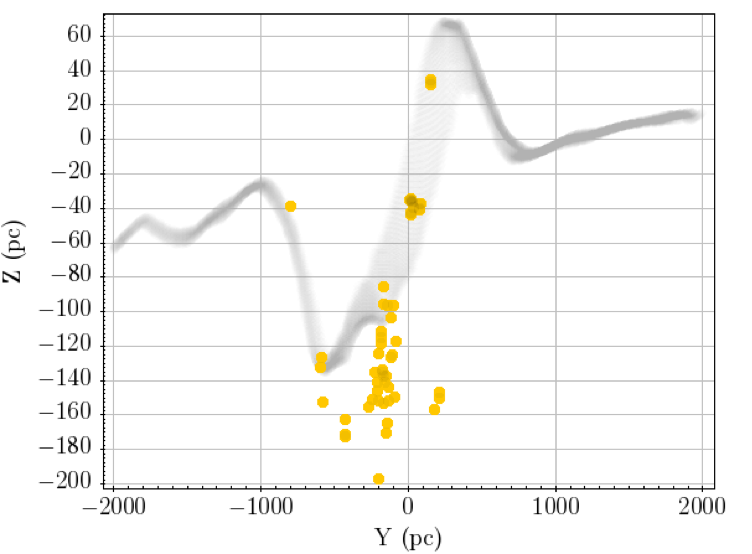}
\caption{Vertical profile, $ Z $ vs. $ Y $, for the Main diagonal of the $ Z(X,Y) $ map (grey band). Molecular gas clouds \citep{Zucker2020} placed along the Main diagonal band are overimposed as yellow full circles. 
YOCs' Kriging map and the molecular clouds show a similar trend,} although the latter seems to show larger displacements with respect to the formal disk in the III Galactic quadrant. This connection between gas and YOCs establishes an important observational constraint for the origin of the phase-space structures.
\label{fig:molecularclouds}
\end{figure}

Thus, we have three well-tested observational features associated with the central box of the solar neighborhood: 1) a highly hierarchical stellar supercomplex of 2.3 kpc in diameter, associated to a  high density of YOCs; 2) the largest $Z$ and $V_Z$ deviations within the entire region under study; and 3) a well-defined linear relationship between $Z$ and $V_Z$. We can add a fourth observational constraint including gas. Molecular gas clouds, catalogued by \citet{Zucker2019, Zucker2020}, show a vertical profile along the Main diagonal ($Y=X$), similar to that found for the YOCs in this work (see Figure \ref{fig:molecularclouds}). Although the $Z$ values of the molecular clouds seem to show a systematic shift with respect to the Main diagonal profile, the agreement between the two distributions is remarkable.  On the other hand, we are not able to discern whether this $Z$-difference between molecular gas and YOCs is real or due to any observational bias, especially considering the different distance-estimation methods for both types of objects \citep[see][for a more detailed explanation of how molecular clouds distances are calculated]{Zucker2019}.  
These four observational issues establish important physical constraints to any proposed scenario to explain the clustered formation and the phase-space structure of the young Galactic disk. 
%

The age of the G-YOC sample is less than 50 Ma, which is too short for the Galactic potential having substantially modified, or erased, the imprint of the stellar cluster formation on the vertical  phase space. 
The vertical perturbations of the disk are typically thought to be originated by different sources injecting momentum and energy into the Galactic disk as mentioned above \citep[e.g.,][and references therein]{Tepper2022}. However, gravitational resonances due to spiral arms, or rotating bars, can also give rise to vertical deviations of the disk, and to ridges and groups in the in-plane velocity components \citep[e.g.,][]{Fux2001, Tatiana2019}, especially if we take into account the Galactic magnetic field \citep{Santillan1999, Gomez_Cox2002, Cox2002, Franco2002}. 

\subsection{Corrugations}
In this section we refer to corrugations in the broadest and most phenomenological sense of the term. That is, as any of the {\sl wavy undulations of the disk components} \citep{Nandakumar2022}. The origin of the term, {\sl vertical corrugation}, is due to \cite{quiroga1974} and so have been named the vertical variations of different disk components (atomic and molecular gas, dust, and stars in different evolutionary states and with different age ranges) in the Milky Way and external galaxies. That includes both azimuthal and radial $Z$ variations in our Galaxy \citep{SF1986}, as well as the wavy profiles observed in gas \citep{uson2008}, dust \citep{Narayan2020} and other wavelengths representative of different stellar populations in edge-on galaxies \citep[][and references therein]{uson2008}. Corrugations are a global disk phenomenon, (wobbly or corrugated disks), but their detection and study has been mainly carried out through the study of 2D profiles along privileged directions (disk profile in edge-on galaxies, spiral arms of the Milky Way, vertical profiles along  Galactocentric radii, etc.) However, we should not forget that both observational \citep{Widrow2012, Romero2019, Pantaleoni-Gonzalez2021} and theoretical \citep{Schonrich2018, Laporte2019} studies converge in considering corrugations a global disk phenomenon. This is our first approach to this concept.

\begin{figure*}
\centering
\includegraphics[width = 0.8\textwidth]{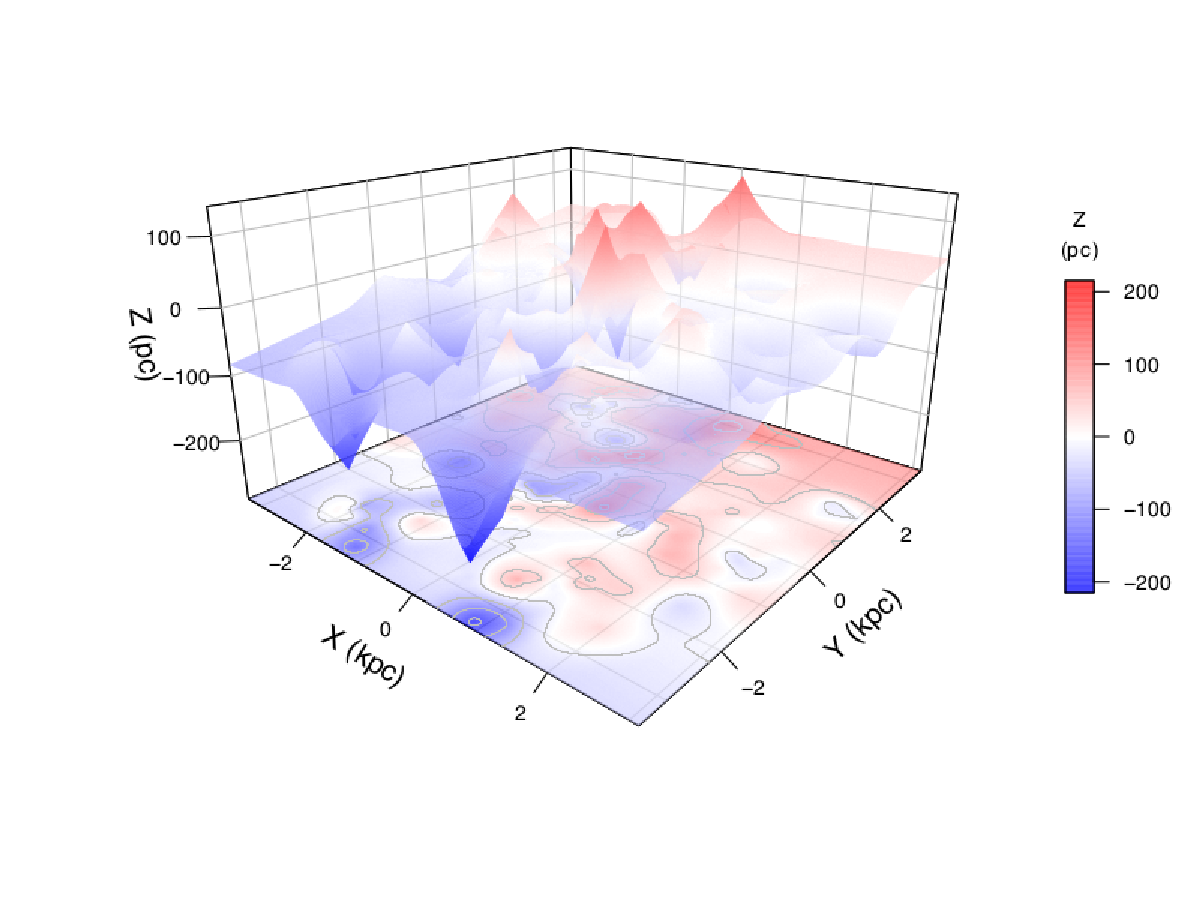}
\includegraphics[width = 0.8\textwidth]{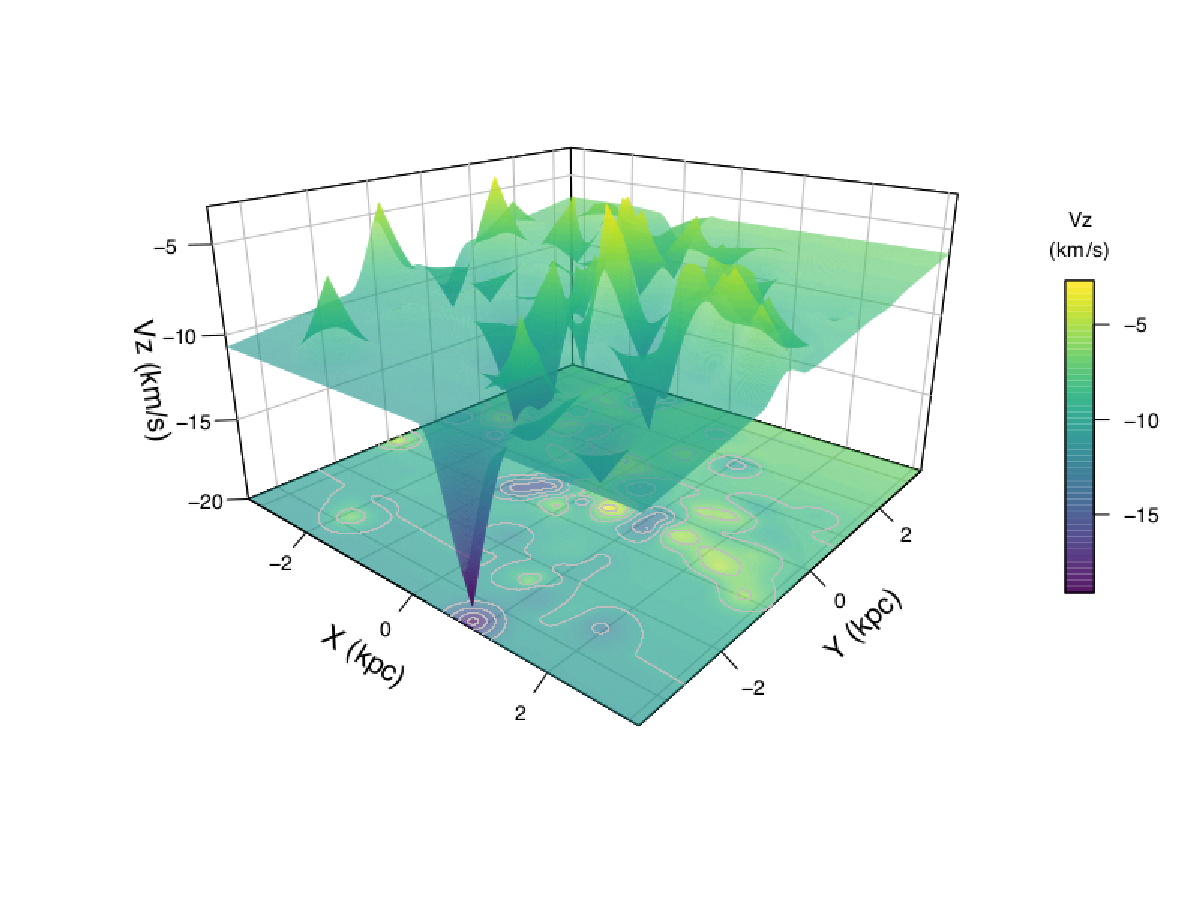}
\caption{3D Topography of the Galactic disk inferred from G-YOC sample.}
\label{fig:corrugations}
\end{figure*}

Our main results are the 3D spatial and vertical-velocity maps of the Galactic disk,  restricted to a radius of 3.5 kpc around the Sun. Figures \ref{fig:KrigZ} and \ref{fig:KrigVz} display a remarkably complex and  high degree structure. We have included a three-dimensional version of them in Figure \ref{fig:corrugations}, for better visualization. 

Spiral galaxy disks display many distinct features, such as spiral arms, central bars,  resonant structures or rings, outer warps \citep[and references therein]{Tepper2022}, inner tilted gas to the Galactic disk \citep{gum1960}, and wave-like disk corrugations \citep[][]{SF1986, AlfaroEfremov1996}.
In particular, spatial corrugations have been known for more than 60 years, but nevertheless their systematic study has been  intermittent in time.  \textit{Gaia} has brought this issue back to the table. Gas vertical-velocity variations have also been detected in some face-on galaxies \citep[][]{alfaro2001, SGAP2015, Nandakumar2022}.  

Figure \ref{fig:linearfeature} already displays these expected wavy-like spatial and kinematic corrugations, throughout different linear features, such as some spiral arms and other selected directions. 
Corrugations are also evident along the Galactocentric radius and azimuthal angle (Figure \ref{fig:warp}), being the $Z$ Galactocentric radius profile  very similar to those found in other edge-on galaxies \citep[e.g.][and references therein]{uson2008}.
The radial vertical-velocity variations quite resemble those corrugated profiles of  $V_Z$ throughout the spiral arms  in nearby spiral galaxies \citep{SGAP2015}.
However, these latter amplitudes  showed  higher values, ranging from 10-15 up to 30-40 \kms in some cases. Regardless, they were obtained from H$\alpha$ data with a completely different observational technique and methodology. 

The maximum amplitude of the corrugations in $Z$ is $31$ pc, and $\approx\,  1$ \kms\  in $V_Z$. The vertical velocity shows a periodogram with wavelengths roughly corresponding to those found spatially, that is between $1.3$ and $2.5$ kpc.
A more exhaustive analysis of the amplitudes and spatial scales of the young Galactic disk  corrugations is beyond the scope of this article. It should be said that the values of the spatial scales found in this work are within the range of values determined in the literature, being the amplitudes slightly lower than those found in other pre\textit{-Gaia}  works \citep{FS1985,SF1986,alfaro1992}.  Moreover, Figure \ref{fig:corrugations} clearly show a kinematically and spatially corrugated disk, and not, necessarily, a collection of wavy-like profiles along some privileged directions.
These plots offer an overview picture of the 3D topography of  YOC's spatial distribution and its perpendicular velocity field. 

The current observations tend to suggest that warps and corrugations are universal phenomena associated with disk galaxies \citep[e.g.][among others]{Quiroga1977, SF1986, florido1991, ACCD1991, uson2008, SGAP2015, Tepper2022}. Despite this universality, what triggers these perturbations has  been barely studied and still poses open questions. Different scenarios have been proposed to explain this phenomenon \citep[see][]{AlfaroEfremov1996,alfaro2001,Alfaro2004,SGAP2015, Nandakumar2022}, in some way linked to mechanisms involved in the large-scale star-formation processes, such as density waves, tidal interactions, collisions with high-velocity clouds, or galactic bore generated from the interaction of a spiral density wave with a thick magnetised gaseous disk, among others \citep{Gomez_Cox2002, Franco2002}.

The hypothesis of an external perturber interacting with the Galaxy, where the main role is attributable  to the passage of Sagittarius dwarf galaxy through the Galactic disk, is gaining more and more strength to explain different vertical-phase features. 
One recent example is found in \cite{Tepper2022}, where they study the effects of a single satellite passing through the Galactic disk. 
They reproduce  stellar and gas corrugations, initially in phase, but disentangling   after a few rotation periods (500-700 Myr). Then, gas and stellar components  evolve in a different way, and their correlation degree could be used  to date this encounter. On the other hand, a damping with time of the corrugation amplitudes is predicted, although the vertical energy of the stars remains almost constant throughout  the galaxy's evolution. 
However, a direct comparison of the complete phase space of the gas and the stellar component is not possible at this time.  We have approached this comparison using the spatial information of  molecular clouds from \citet{Zucker2019, Zucker2020}, shown in Figure \ref{fig:molecularclouds}. 

Systematic studies of the gas and stellar population in different age ranges, would help understand these structures,  their possible origin, or whether these phenomena arise from external action rather than from internal processes. 

\subsection{Scenario}

The 4D view, ($X,\, Y,\, Z,\, V_Z$), of the central box ($2\times2$ kpc) obtained in this work (Figures \ref{fig:KrigZ} and \ref{fig:KrigVz}) introduces important constraints in proposing a global scenario to explain the intricate pattern of the vertical phase space. The main conditionant stems from the narrow age range of the $Gaia-$YOC sample ($\leq  10^{7.5}$ a). Otherwise, the observed almost-linear relationship between $V_Z$ and $Z$ is different from the spiral shapes found for older stellar populations covering a wider age range \citep{2018Natur.561..360A, Bland_Galah2019, Li2021}. The steep local gradients in both $Z$ and $V_Z$, together with the youth of the tracers, would suggest a transient phenomenon to produce large local deformations of the vertical phase. The relative position of molecular gas and YOCs in the solar neighborhood (Figure \ref{fig:molecularclouds}) suggests that this encounter is recent. Although the gas and stars could have been in phase for much longer periods ($\sim$ 500 Ma), it is more difficult to explain that the maximum deviation from the plane, with amplitude far in excess of the secondary corrugations, was not damped in that time period \citep[][]{Tepper2022}. The correlation between $Z$ and $V_Z$ (Figure \ref{fig:central_box_phase}), the larger amplitudes at  (Figure \ref{fig:linearfeatures3}), and the evident spatial connection between molecular gas and young stars (Figure \ref{fig:molecularclouds}) suggest that the encounter has taken place here and now. There are indications that the leading arm of the Sagittarius dwarf galaxy may be entering the Galactic disk in the solar neighborhood \citep{Martinez_Delgado2007}. The quasi-linear relationship between $Z$ and $V_Z$ for a perturbed gaseous disk is inferred from the collision between HVCs and a magnetized disk \citep{Santillan1999}.

The behavior of the older stellar population is out of our study, and we can say nothing about that, so far. However, the perturbation of the Galactic disk by encounters with momentum and energy injectors of different nature, could account for the observed gas and YOCs features, as well as for the generation of vertical phase spiral when the old disk population is taken into account \citep{2018Natur.561..360A}.
The high degree of hierarchy in the distribution of YOCs and the existence of several moving groups in the central region suggest that a dynamic trap associated with gravitational resonances could  explain this phenomenology \citep[][]{Lepine2018, Tatiana2019, Barros2020}. 

If we move to larger spatial scales, we observe vertical behaviors  that have  been traditionally associated with, and have supported, the presence of Galactic warping. However, the main feature we observe is a gradient in $V_Z$ with $Y$ (see Figures \ref{fig:KrigVz}, \ref{fig:warp_Z_Vz_Y}) that takes an average value of $0.5-0.6$ \kmskpc. 
Comparing the $V_Z(X, Y)$ map of YOCs with that obtained for the OB-sample's proper motions in latitude of \cite{Romero2019} for the same region, we observe a similar pattern. This structure is no longer detected in the RGB sample of these authors. On the other hand, different simulations of the behavior of $\mu_b$ for different warping shapes \citep{Romero2019} foresee a range of variations four times larger than observed in that work, and its predicted spatial distribution does not fit  the observed vertical velocity gradient in the solar neighborhood. All this leads us to consider that, at least, for the young stellar population, transient phenomena dominate the structure of the velocity map, and classical warping approaches do not fit detailed observational kinematic maps of the younger stellar population in the solar neighborhood. 

Thus, gravitational resonances and frequent encounters with passing perturbers appear to be the main drivers of the complex structure of the phase subspace analyzed in this paper. The large vertical deviations in both $Z$ and $V_Z$ in the solar neighborhood, the quasi-linear relationship between both variables, and the spatial connection between molecular gas and YOCs lead us to conclude that such an encounter has taken place not longer than 50 Ma ago.

\section{Conclusions} 
\label{sec:conclusions}

In this work, we have applied the Kriging inference technique to a sample of \textit{Gaia} Young Open Clusters (G-YOCs) in order to study the 3D spatial structure and the vertical velocity field of the Galactic disk around the Sun. In general, the sample of G-YOC exhibits a clumpy, hierarchical distribution with a 3D fractal dimension in the range $D_f = 2.5-2.6$, similar to those values obtained for different molecular clouds in the Milky Way. This supports the idea that young stars and clusters distribute following the spatial structure of the interstellar medium from which they were formed. A detailed analysis of this hierarchical structure yields a set of groupings embedded within other groupings that can be identified with the so-called stellar complexes or supercomplexes. Our results reveal a complex structure that is far from a simple planar geometry, in which warp, corrugations, and large spatial and kinematic deviations at the solar neighborhood, are part of a single but complex picture. There is a large-scale spatial variation in both the 3D spatial distribution and the vertical-velocity field  associated with the known Galactic warp. There is also an intermediate-scale ($2 \times 2$ kpc) region with a high degree of spatial and kinematic structure, including the so-called Gould Belt,  delimited in $Z$ by the positions of the Ori OB and Sco-Cen OB associations. Superimposed throughout the entire structure, small-scaled corrugations are observed both in the spatial distribution and in the vertical velocity field of the clusters, with peaks and valleys of different amplitudes and scales. These corrugations affect the entire disk and, in general, can be observed in any direction on the plane. The most pronounced of these fluctuations is located close to the  Sun's position, where the $V_Z (X,Y)$ and $Z (X,Y)$ fields show a clear correlation, which we associate with a transient phenomenon produced by a relatively fast transfer of energy and momentum into a magnetized gas disk. We discussed the physical mechanisms  behind this complex structure, but, in any case, our results provide a global 4D ($X,Y,Z,V_Z$) picture that imposes important constraints on models of Galactic disk formation and evolution.

From this analysis we conclude that:
\begin{enumerate}
    \item The existence of a hierarchy of star formation with at least three levels, corresponding in general to the complexes and supercomplexes detected by \citetalias{EfremovSitnik88}.
    \item Supercomplex A, the highest hierarchy level, is located in the immediate solar vicinity with a maximum diameter of 2.3 kpc. It contains most of the region's moving groups, and its origin and existence could be explained by dynamic traps associated with Galactic resonances.
    \item We observe three different behaviors in the structure of the maps $Z(X, \, Y)$ and $V_Z(X, \, Y)$:  (i) at large scales, we notice a gradient of both variables with the $Y$ coordinate; (ii) the central region shows the largest deviations of $Z$ and $V_Z$, and a linear relationship between both variables; and (iii) superimposed on these features, we see peaks and valleys in both maps with varying (but low) amplitude and spatial scales between $1. 3$ and $2.5$ kpc.

    \item  The combined action of encounters with passing perturbers and the resonances generated by an asymmetric and non-stationary Galactic potential can explain, at least qualitatively, these observations. 
    \item We consider that, on a short spatial scale, we are observing the effects of the last passage of the Sagittarius dwarf galaxy  no more than 50 Ma ago.
\end{enumerate}




\acknowledgments
We are grateful to Bruce Elmegreen for comments on the manuscript. This work was motivated in memoriam to our friend and colleague Yuri N. Efremov. Bruce Elmegreen and Emilio J. Alfaro are forever grateful to him for his inspiring insight into the hierarchical structure of star formation through his cataloging of star complexes and supercomplexes in the Milky Way and neighboring galaxies. He brought us together in 1992 on the island of Elba and we were friends and collaborators until his untimely passing in 2019. We acknowledge the referee for a careful reading of the manuscript and very useful comments, which helped us to improve the paper. Emilio J. Alfaro and Manuel Jim\'enez acknowledge financial support from the State Agency for Research of the Spanish MCIU through the ``Center of Excellence Severo Ochoa'' award to the Instituto de Astrof\'isica de Andaluc\'ia (SEV-2017-0709). Emilio J. Alfaro and Carmen Sánchez-Gil have been supported by PY20-00753 grant from Junta de Andalucía, Spain (Autonomic Government of Andalusia). Néstor Sánchez and Jesús Maíz-Apellániz acknowledge financial support from grants PGC2018-095049-B-C21 and PGC2018-095049-B-C22 (MCIU), respectively. In this work, we have made extensive use of TopCat \citep{Taylor2005}. We thank its author, and subsequent contributors, for the creation and development of this tool.

\appendix 

\renewcommand\thefigure{\thesection.\arabic{figure}} 
\setcounter{figure}{0}

\renewcommand\thetable{\thesection.\arabic{table}} 
\setcounter{table}{0}

\section{Variogram Model Selection}\label{App:VMS}

\subsection{Variogram Estimation and Model Fitting}

\begin{figure*}[t]
    \centering
    \includegraphics[width=0.95\textwidth]{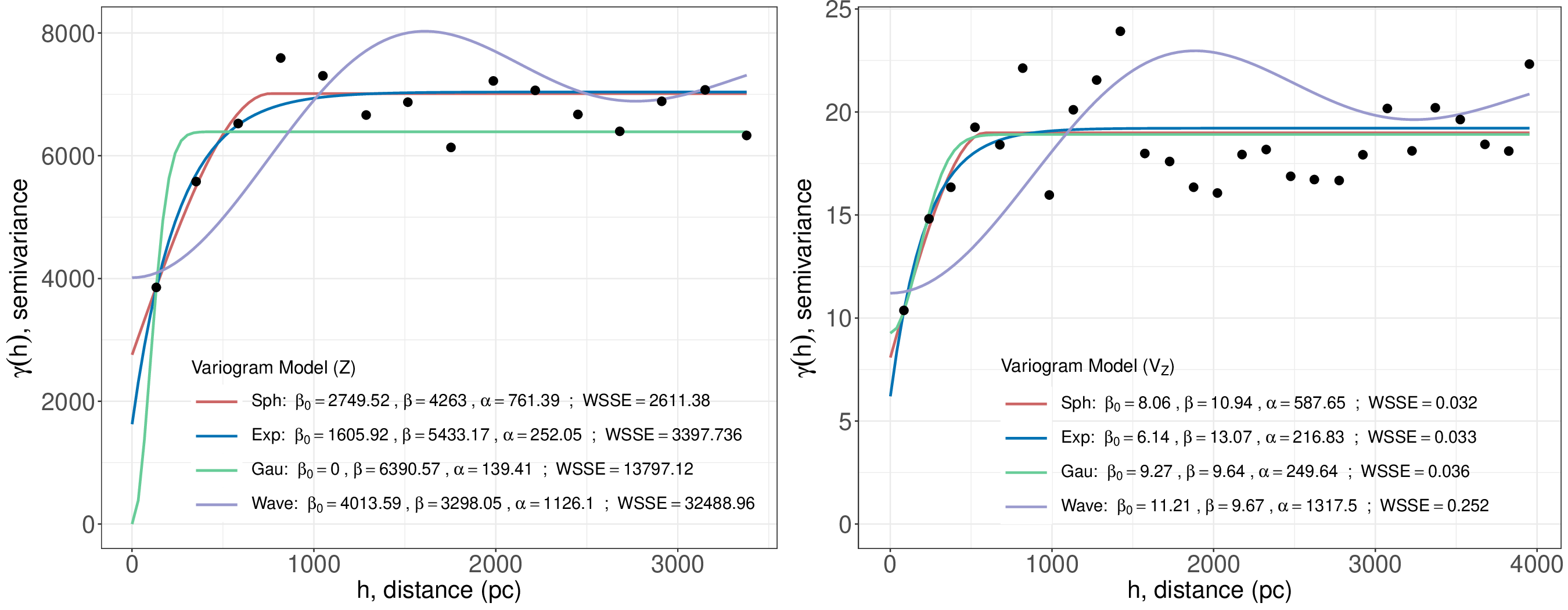}
    \caption{Variogram fit model selection, corresponding to the $Z$ Galactic coordinate on the left, and for $V_Z$ on the right. It has been tested four different models for the Variogram fit: spherical `Sph', exponential `Exp', Gaussian `Gau', and wave. The spherical model appears to be the best choice, having the lowest weighted sum of errors (WSSE) statistics.}
    \label{fig:variograms_all}
\end{figure*}

Spatial prediction refers to the prediction of the unknown quantity $ Z(r_0) $, based on sample data $\left\{Z(r_i)\right\}_{i=1,\ldots,n}$ and assumptions regarding the form of the trend of $Z$ and its variance and spatial correlation. 
For Universal Kriging model, we assume $Z(r) = m(r) + e(r)$ with $E(e(r))=0$, and the mean, which varies spatially, is modeled as a linear function of $p$ known predictors $F = (f_1(r),\ldots,f_p(r))$, with $f_j(r) = (f_j(r_1),\ldots,f_j(r_n))^T$. This is 
\begin{equation}
    Z(r) = m(r) + e(r) = {F}\beta + e(r)
\end{equation} 
with unknown regression coefficients $\mathbf{\beta} =   (\beta_1,\ldots,\beta_p)^T$ (see Equation \eqref{eq:krig_model}). 
Given this model, the best linear unbiased predictor, or the Kriging predictor, of $Z(r)$ consists of an estimated mean value for location $r_0$, plus a weighted mean of the residuals from the mean function, with weights $v^T V^{-1}$ known as the simple Kriging weights, 
\begin{equation}
\label{eq:krig_pred}
\hat Z(r_0) = f(r_0)\hat \beta  + v^TV^{-1}(z(r)-F\hat\beta)
\end{equation}
where $f(r_0) = (f_1(r_0),\ldots,f_p(r_0))$, $z(r) = (Z(r_1),\ldots,Z(r_n))^T$,
$v = \big(Cov(e(r_1),e(r_0)),\ldots,Cov(e(r_1),e(r_n))\big)$ the covariance vector of $Z(r_0)$ and $z(r)$,
$V = Cov(e(r))$ the known covariance matrix of $Z(r)$, 
and $\hat \beta = (F^T V^{-1} F)^{-1}F^T V^{-1}z(r)$ the generalized least squares estimate of the trend coefficients $\beta$. 
The Kriging prediction has prediction variance, or Kriging variance, given as 
\begin{eqnarray}
    \label{eq:sigma_krig}
    \sigma^2_{Z}(r_0) = Var(Z(r_0)-\hat{Z}(r_0)) &=&  
    Var(Z(r_0)) - v^TV^{-1}v +  \\
    && \left(f(r_0) - v^TV^{-1}F\right) (F^T V^{-1} F)^{-1} \left(f(r_0) - v^TV^{-1}F\right)^T \nonumber
\end{eqnarray}
The variogram estimation is a significant issue for statistical inference of spatially correlated variables. Typically, its estimation is split into two stages: \textit{empirical variogram estimation} $\hat{\gamma}(h)$, 
and \textit{model fitting} with a valid theoretical model ${\gamma}(h)$ to capture the actual spatial dependence of the data. Variogram analysis provides a helpful tool for summarizing and measuring spatial dependence in spatial data.  
However, its main contribution is to estimate the value of the spatial variable at an unsampled location within some inference procedures, such as the Kriging method. The latter approach differs from classical regression in that local features can affect the solution, and so it considers a weighted mean. 
Therefore estimating a variogram plays a decisive role, as it is commonly used to find the optimal values of the weights. 

We considered the widely used exponential and the spherical semivariogram models, as well as the Gaussian and the wave models. The latter was also considered because of its atypical irregular behaviour.
\begin{eqnarray}
    \gamma_{sph}(h) \ \ &= & \ \ \begin{cases}
    \beta_0 + \beta \left( \frac{3h}{2\alpha}- \frac{1}{2}\left(\frac{h}{\alpha} \right)^3\right),
    & 0 < h \leq \alpha\\
    \beta_0 + \beta & h > \alpha 
    \end{cases} 
    \label{Eq:Sph_VM}\\[1ex]
    \gamma_{exp}(h) \ \ &= & \ \ 
    \beta_0 + \beta \left( 1 - e^{-\frac{h}{\alpha}}\right), 
    \quad h > 0 
    \label{Eq:Exp_VM}\\[1ex]
    \gamma_{gau}(h) \ \ &= & \ \
    \beta_0 + \beta \left( 1 - e^{-\frac{h^2}{\alpha^2}}\right), 
    \quad h > 0 
    \label{Eq:Gau_VM}\\[1ex]
    \gamma_{wave}(h) \ \ &= & \ \
    \beta_0 + \beta \left( 1 - \alpha \, \frac{sin\left(-{h}/{\alpha}\right)}{h} \right), 
    \quad h > 0
    \label{Eq:Wave_VM}
\end{eqnarray}
The main characteristics, or parameters, of many semivariogram models are: {\it (i)} the {\it nugget} effect, $\beta_0$. In theory, the semivariogram should be zero for a lag distance $h$ of zero, but in practice it can be significantly different from zero. This non-zero value $\beta_0$ reflects local effects or sampling error; 
{\it (ii)} Another important characteristic is that the semi-variance function $\gamma(h)$ can approach, or asymptotically converge to, a constant value known as the {\it sill}, $\beta_0 + \beta$. When there is a nugget $\beta_0 > 0$, $ \beta$ is called {\it partial sill};
{\it (iii)} The distance at which the semivariogram reaches the sill is called the {\it range}, $\alpha$. It means that there is a distance beyond which the correlation between variables is zero. Sample locations separated by distances shorter than the range are spatially auto-correlated, while locations that are further apart than the range are not.

Figure \ref{fig:variograms_all} shows the different variogram models described above, for both the $Z$ and $V_Z$ variables, fitted to the sample or empirical semi-variogram $\hat \gamma (h)$, represented with black dots. 
The sample variograms and co-variograms are calculated from predicted residuals 
$\hat{e}(r_i) = z(r_i) - \hat{m}(r_i)$, 
with $\hat{m} (r_i) = F\hat{\beta}$ the ordinary least square estimates of the mean, as (equivalent to Equation \eqref{eq:variogram})
\begin{equation}
\label{eq:variogram_bis}
\hat \gamma(\bar{h}_j) = \frac{1}{2N_j} \displaystyle \sum_{i = 1}^{N_j} \left( \hat{e}({r}_{i}) - \hat{e}({r}_{i} + {h}) \right)^{2}, \quad \forall (r_i,r_i+h) : h \in [h_j,h_j+\delta]
\end{equation}
this is for a number $N_j$ of pairs within a regular distance interval $[h_j,h_j+\delta]$, with $\bar{h}_j$ the average of such interval.  

\subsection{Cross-Validation and Model Selection}

\begin{table}[t]
\caption{Variogram fit results.}
    \centering
    \begin{tabular}{r ccc ccc}
       \hline
      \multirow{2}{*}{Model}  &
      \multicolumn{3}{c}{$Z$} & \multicolumn{3}{c}{$V_Z$} \\
       \cmidrule(lr){2-4}\cmidrule(lr){5-7} 
         &WSSE& AIC & $R^2$ & WSSE&AIC & $R^2$ 
         \\[0.5ex] 
         \hline
        Exponential  & 3397.74   &   -257.89  &   99.98 & 
        0.03 & -151.36 &    93.54  \\[0.5ex]
        Gaussian  & 13797.12 & -257.86  &   98.34 & 
        0.04 &   -116.22 &    93.58 \\[0.5ex]
        Spherical  & 2611.38 &   -257.89  &   99.98 & 
        0.03 &   -150.99 &    93.35 \\[0.5ex]
        Wave  &  32488.96  &  -257.98  &  99.98 &  
        0.25  &   -152.53  &   92.5 \\\hline 
    \end{tabular}
    \label{tab:Table_AIC_WSSE}
\end{table}

The fitting method at the \textit{model variogram fitting} stage uses non-linear regression to fit the coefficients or parameters of the model. 
For this, it is  minimised the following weighted sum of square errors, 
\begin{equation}
    \label{eq:WSSE}
    WSSE = \sum_{j=1}^{n} \omega_j \left( \hat{\gamma}(\bar{h}_j) - \gamma(\bar{h}_j)\right)^2
\end{equation}
where $\gamma(h)$ is according the chosen variogram model (from \eqref{Eq:Sph_VM} to \eqref{Eq:Wave_VM}), $\hat{\gamma}(h)$ is the empirical variogram estimation \eqref{eq:variogram_bis}. We have chosen the weights to be the ratio between the number of data point pairs grouped into the corresponding distance intervals and the distance itself, i.e. $\omega_j = N_j / h_j$. 

We used this statistic to evaluate how well the different models fit the data and determine which one is the best fit for the data. We also compute the Akaike information criterion (AIC) to compare the different possible models, however the AIC results were pretty similar in all cases (bear in mind that all these models have the same number of parameters). The WSSE results to be more discriminant for the case of the $Z$ component, as can be checked in Figure \ref{fig:variograms_all}, where the spherical model seems to fit better the data. For the case of the $V_Z$ component, the results were also similar for both WSSE and AIC between the different models. Therefore, we choose the spherical model for variogram fitting based on the discriminant result in $Z$. 

Besides the above statistics criterion, we also carry on cross-validation, another model diagnostics for testing the quality of the variogram fitting itself and model comparison. 
The data set is split into two sets: a modelling set and a validation one (we took $1/5$ of the data or a $n=5$-fold partition). The former is used for variogram modelling and Kriging on the locations of the validation set. Then the validation measurements can be compared to their predictions. 
This process is repeated a hundred times, choosing the modelling and validation sets randomly at each run. 

We can then compute the sum of squared errors from the mean or $R^2$, another goodness of the fit measurement, which indicates how much Kriging prediction is a better predictor than the mean. The average $R^2$ is around the $99\%$ for all models in the $Z$ component (slightly lower for the Gaussian model), and around the $93\%$ for $V_Z$ variable. 
So $R^2$ is not a determinant for model selection in this case, see Table \ref{tab:Table_AIC_WSSE}. 

\begin{figure*}
    \centering
     \includegraphics[width=0.75\textwidth]{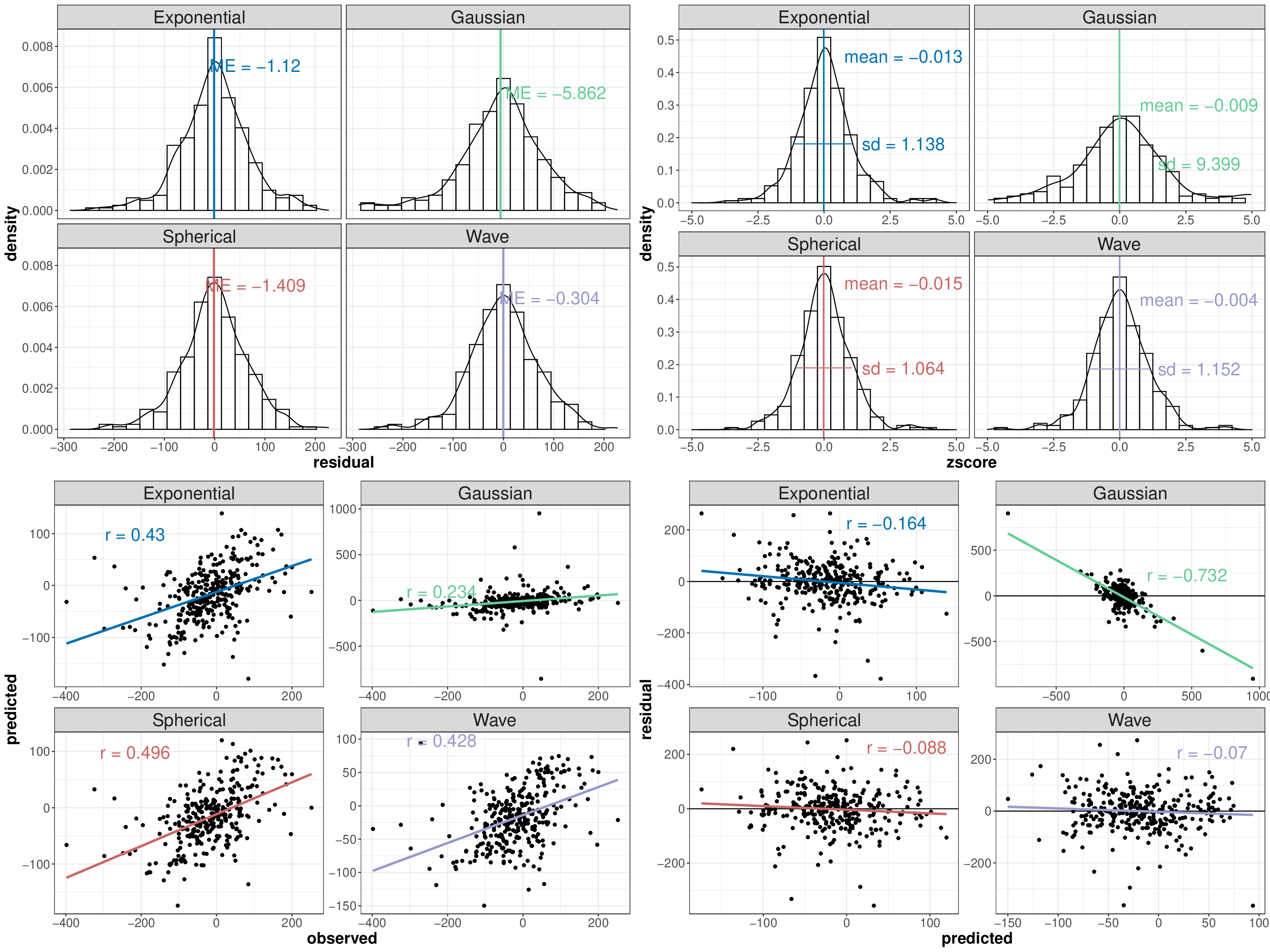} 
    \caption{Cross-validation analysis for $Z$ Variogram model selection.
    On the bottom left panels, correlation between observed and predicted data, ideally close to 1. 
    On the bottom right panels, correlation between predicted and residuals, ideally close to zero. In both cases, except for the Gaussian model, the results are quite similar.}
    \label{fig:CV_Z}
\end{figure*}

\begin{figure*}
    \centering
    \includegraphics[width=0.75\textwidth]{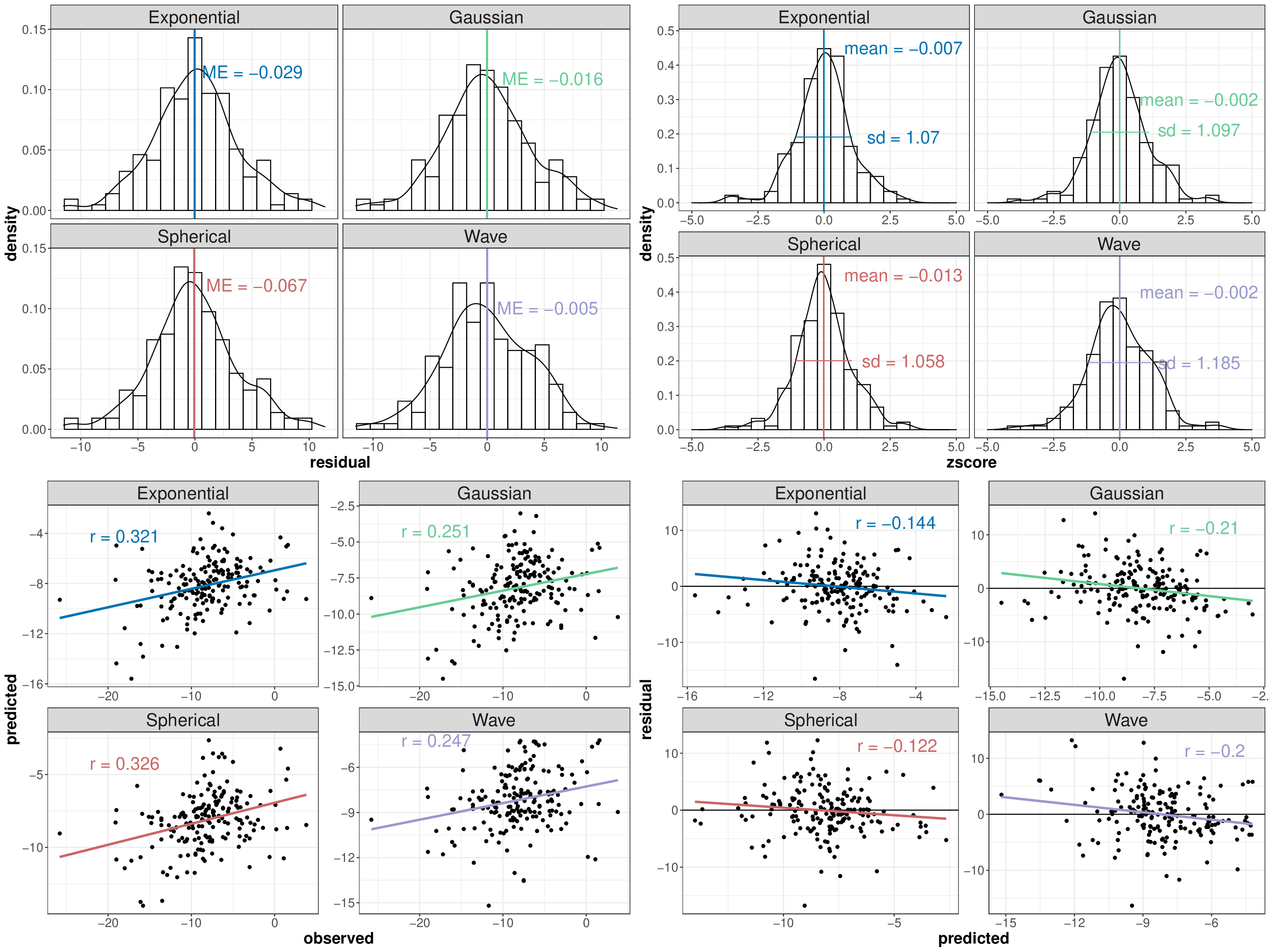} 
    \caption{Analogously to Figure \ref{fig:CV_Z}, cross-validation analysis for $V_Z$ Variogram model selection.}
    \label{fig:CV_VZ}
\end{figure*}

We also inspect the errors from the cross-validation analysis, such as the residuals $Z(r_i) - \hat{Z}(r_i)$, and z-scores $z_i = (Z(r_i) - \hat{Z}(r_i)) / \sigma_{krig}(r_i)$ (the standardized residual but taking into account the corresponding Kriging standard error).
Figures \ref{fig:CV_Z} and \ref{fig:CV_VZ} display these cross-validation errors on the top panels, for one of the random runs. 
If the variogram fitting is correct, cross-validation residuals are small, have zero mean, and have no apparent structure. This appears to be the case for the four models, for both $Z$ and $V_Z$.  
In contrast to standard residuals, the z-score takes the Kriging variance into account. Whereas we expect small residuals with no trend, if the variogram model is correct, the z-score should have mean and variance values close to 0 and 1, respectively. 
For the $Z$ variable, we can discard the Gaussian model for having a dispersion quite further than 1, which may imply some misfitting. The Spherical model would head the model ranking but with little difference from the Exponential one.
For the $V_Z$ variable, we cannot see meaningful differences between the models.   

Finally, we check how these errors are related with respect to the predictions $\hat{Z}(r_i)$ and the observed $Z(r_i)$, or test data set at each $n$-fold, at the bottom panels of Figures \ref{fig:CV_Z} and \ref{fig:CV_VZ}. 
A visual verification that the variogram model fit is sensible has been checked through the observations vs. predictions scatter plots (left bottom panels) and predictions vs. residuals plots (right bottom panels). We expect a high correlation (the higher correlation, the better model) for the former and a low correlation for the latter (the residuals should be random without any trend). 
We can observe that the Spherical and Exponential models are slightly better than the other two. The different models yield similar results for the $V_Z$ variable, where the Spherical and Exponential models are somewhat better, analogously to the $Z$ case. The only exception is the Gaussian model for the $Z$ variable, where the predictions versus residuals' plot show a clear trend likely attributed to a misfitting. 
Then, we chose the Spherical model based on this final model diagnosis jointly with a moderately better WSSE statistic. We used the same model for homogeneity for both $Z$ and $V_Z$ variables. 


\clearpage
\bibliography{Manuscript}{}
\bibliographystyle{aasjournal}

\end{document}